\documentstyle[12pt,epsf,epsfig]{article}
\newfont{\thiplo}{msbm10 scaled\magstep 2}
\newfont{\gothic}{eufb10 scaled\magstep 2}
\newfont{\unc}{eurb10} 
\newskip\humongous \humongous=0pt plus 1000pt minus 1000pt
\def\caja{\mathsurround=0pt}
\def\eqalign#1{\,\vcenter{\openup1\jot \caja
        \ialign{\strut \hfil$\displaystyle{##}$&$
        \displaystyle{{}##}$\hfil\crcr#1\crcr}}\,}
\newif\ifdtup
\def\panorama{\global\dtuptrue \openup1\jot \caja
        \everycr{\noalign{\ifdtup \global\dtupfalse
        \vskip-\lineskiplimit \vskip\normallineskiplimit
        \else \penalty\interdisplaylinepenalty \fi}}}
\def\eqalignno#1{\panorama \tabskip=\humongous 
        \halign to\displaywidth{\hfil$\displaystyle{##}$
        \tabskip=0pt&$\displaystyle{{}##}$\hfil 
        \tabskip=\humongous&\llap{$##$}\tabskip=0pt
        \crcr#1\crcr}}

\def\eqright #1\cr{\noalign{\hfill$\displaystyle{{}#1}$}}
\def\eqleft #1\cr{\noalign{\noindent$\displaystyle{{}#1}$\hfill}}

\def\oldreffmt#1{\rlap{[#1]} \hbox to 2\parindent{}}

\def\figfmt#1{\rlap{Figure {#1}} \hbox to 1in{}}

\def\sectioneq{\def\theequation{\thesection.\arabic{equation}}{\let
\holdsection=\section\def\section{\setcounter{equation}{0}\holdsection}}}%

\newcounter{holdequation}

\def\num{(\refstepcounter{equation}\theequation)}
\def\auto{\eqno(\refstepcounter{equation}\theequation)}
\def\begineq #1\endeq{$$ \refstepcounter{equation}\eqalign{#1}\eqno
	(\theequation) $$}
\def\contlimit{\,{\hbox{$\longrightarrow$}\kern-1.8em\lower1ex
\hbox{${\scriptstyle (a\rightarrow0)}$}}\,}
\def\centeron#1#2{{\setbox0=\hbox{#1}\setbox1=\hbox{#2}\ifdim
\wd1>\wd0\kern.5\wd1\kern-.5\wd0\fi
\copy0\kern-.5\wd0\kern-.5\wd1\copy1\ifdim\wd0>\wd1
\kern.5\wd0\kern-.5\wd1\fi}}
\def\centerover#1#2{\centeron{#1}{\setbox0=\hbox{#1}\setbox
1=\hbox{#2}\raise\ht0\hbox{\raise\dp1\hbox{\copy1}}}}
\def\centerunder#1#2{\centeron{#1}{\setbox0=\hbox{#1}\setbox
1=\hbox{#2}\lower\dp0\hbox{\lower\ht1\hbox{\copy1}}}}
\def\lsim{\;\centeron{\raise.35ex\hbox{$<$}}{\lower.65ex\hbox
{$\sim$}}\;}
\def\gsim{\;\centeron{\raise.35ex\hbox{$>$}}{\lower.65ex\hbox
{$\sim$}}\;}

\def\super#1{\ifmmode \hbox{\textsuper{#1}}\else\textsuper{#1}\fi}
\def\textsuper#1{\newcount\holdspacefactor\holdspacefactor=\spacefactor
$^{#1}$\spacefactor=\holdspacefactor}

\def\getcite#1,{\advance\citenumber by1
\def\getcitearg{#1}\def\lastarg{@}
\ifnum\citenumber=1
\ref{#1}\let\next=\getcite\else\ifx\getcitearg\lastarg\let\next=\relax
\else ,\ref{#1}\let\next=\getcite\fi\fi\next}

\def\pom{{\rm P\kern -0.53em\llap I\,}}
\def\spom{{\rm P\kern -0.36em\llap \small I\,}}
\def\sspom{{\rm P\kern -0.33em\llap \footnotesize I\,}}

\relax
\def\contlimit{\,{\hbox{$\longrightarrow$}\kern-1.8em\lower1ex
\hbox{${\scriptstyle (a\rightarrow0)}$}}\,}
\def\upon #1/#2 {{\textstyle{#1\over #2}}}
\relax
\renewcommand{\thefootnote}{\fnsymbol{footnote}}

\sectioneq

\def\til#1{\centeron{\hbox{$#1$}}{\lower 2ex\hbox{$\char'176$}}}
\def\tild#1{\centeron{\hbox{$\,#1$}}{\lower 2.5ex\hbox{$\char'176$}}}
\def\sumtil{\centeron{\hbox{$\displaystyle\sum$}}{\lower
-1.5ex\hbox{$\widetilde{\phantom{xx}}$}}}

\def\kbar{\underline{k}}

\newcommand{\bit}{\begin{itemize}}
\newcommand{\eit}{\end{itemize}}

\newcommand{\beq}{\begin{equation}}
\newcommand{\eeq}{\end{equation}}
\newcommand{\beqa}{\begin{eqnarray}}
\newcommand{\eeqa}{\end{eqnarray}}

\begin{document} 

\begin{titlepage} 

\rightline{\vbox{\halign{&#\hfil\cr
&ANL-HEP-PR-00-011\cr
&\today\cr}}} 
\vspace{1.25in} 

\begin{center} 
 
{\large\bf 
THE PAST AND FUTURE OF S-MATRIX THEORY }\footnote{Work 
supported by the U.S.
Department of Energy, Division of High Energy Physics, \newline Contracts
W-31-109-ENG-38 and DEFG05-86-ER-40272} 
\medskip

Alan. R. White\footnote{arw@hep.anl.gov }

\vskip 0.6cm

\centerline{High Energy Physics Division}
\centerline{Argonne National Laboratory}
\centerline{9700 South Cass, Il 60439, USA.}
\vspace{0.5cm}

\vspace{2in}

Contribution to ``Scattering'', edited by E.~R.~Pike and P.~Sabatier, 

to be published by Academic Press.

\end{center}

\renewcommand{\thefootnote}{\arabic{footnote}} \end{titlepage}

\setcounter{page}{1}

~
\vspace{0.5in}

{\hoffset 1.5in 

{\bf CONTENTS}
}

\vspace{0.5in}

\begin{itemize}
\item[{1.}]{\bf Introduction.} 
\item[{2.}]{\bf The Early Years. } 
\begin{itemize}
\item[{2.1}]{Living Without Field Theory. }
\item[{2.2}]{Initial Postulates.}
\item[{2.3}]{Further Postulates.}
\item[{2.4}]{QCD and a Final Postulate.}
\item[{2.5}]{Analyticity in Field Theory.}
\end{itemize}
\item[{3.}]{\bf Axiomatic S-Matrix Theory. }
\begin{itemize} 
\item[{3.1}]{Unitarity, Bubble Diagrams and Landau Diagrams. }
\item[{3.2}]{The Landau Equations and the ``$+~ \alpha$'' Condition.} 
\item[{3.3}]{Macrocausality and Essential Support.}
\item[{3.4}]{The Structure Theorem.}
\item[{3.5}]{Local Discontinuity Formulae.}
\item[{3.6}]{Good and Bad Functions and the Steinmann Relations.} 
\item[{3.7}]{Global Discontinuity Formulae.}
\item[{3.8}]{CPT, Hermitian Analyticity, etc.}
\item[{3.9}]{Holonomy.}
\end{itemize}
\item[{4.}]{\bf Asymptotic S-Matrix Theory. }
\begin{itemize}
\item[{4.1}]{The Elastic Scattering Asymptotic Dispersion Relation.}
\item[{4.2}]{Multiparticle Kinematics and Analyticity Domains.}
\item[{4.3}]{Multiparticle Asymptotic Dispersion Relations. }
\item[{4.4}]{Classification of Multiple Discontinuities.}
\end{itemize}
\end{itemize}
\newpage

~
\vspace{0.5in}
{\hoffset 0.5in 
\begin{itemize}
\item[{5.}]{\bf Multi-Regge Theory.}
\begin{itemize}
\item[{5.1}]{Partial-Wave Expansions.}
\item[{5.2}]{Froissart-Gribov Continuations.}
\item[{5.3}]{Sommerfeld-Watson Representations. }
\item[{5.4}]{Reggeon Unitarity. }
\end{itemize}
\item[{6.}]{\bf Reggeon Field Theory.}
\begin{itemize}
\item[{6.1}]{Pomeron Phase-Space and the Effective Lagrangian.}
\item[{6.2}]{The Critical Pomeron.}
\item[{6.3}]{The Super-Critical Pomeron.}
\end{itemize}
\item[{7.}]{\bf QCD and the Critical Pomeron.}
\begin{itemize}
\item[{7.1}]{Reggeon Diagrams in QCD. }
\item[{7.2}]{Color Superconductivity and the Super-Critical Pomeron.}
\item[{7.3}]{Quark Saturation and an Infra-Red Fixed Point.}
\item[{7.4}]{Uniqueness of the S-Matrix ? } 
\end{itemize}
\end{itemize}
}
\newpage

\section{Introduction.}

S-Matrix Theory was initially developed as a means to go beyond the
weak-coupling perturbation expansions of field theory and at the same time
avoid infinite renormalization effects. Landau was the first to
emphasize that the singular or ``imaginary'' parts of perturbative diagrams
involve only on mass-shell intermediate states and so are free of 
renormalization problems
(Landau, 1959). He also argued that the analyticity properties needed to
reconstruct a full amplitude from the imaginary part, via a dispersion
relation, would be a consequence of the underlying local causality of 
the interaction. The arguments of Landau, and many others, 
raised the hope that if analyticity properties were
exploited a strong interaction S-Matrix could be self-consistently 
calculated diagrammatically without encountering divergences, even when 
this interaction could not be described by field theory. (In fact the
absence of a Mandelstam representation (Mandelstam, 1958,1961) for multiparticle
amplitudes prevents an iterative perturbation expansion
using discontinuities plus dispersion relations - a problem that is
currently circumvented only in the regge limit.) At the time it was 
believed that local quantum field
theory could not produce an interaction for which multiple short-distance
interactions did not become uncontrollably strong\footnote{We use
``uncontrollable'' as a non-technical description of the technical property
that the perturbation expansion in powers of the coupling constant is so
wildly divergent that there is no summation procedure that allows it to be
used to define the theory.} (Landau 1955; Landau and Pomeranchuk 1955). 
The short-distance decrease of the
coupling (asymptotic freedom) in non-abelian gauge theories such
as QCD (Gross and Wilczek, 1973,1974; Politzer, 1974)
was not known, of course. In a sense, the attempts to develop
S-Matrix Theory as an independent formalism based on analyticity anticipated
that the strong interaction would eventually be understood as a local
interaction. QCD was formulated over twenty-five years ago and is now
established\footnote{The detailed formulation of the Standard Model,
including both Quantum Chromodynamics (QCD) and the Electroweak 
Interaction, is described in later chapters.}  as the (underlying) field
theory of the strong interaction. Paradoxically, perhaps, the locality of
the interaction is described in terms of gauge-dependent fields whose
relation to the asymptotic states of the theory is so complicated that it is
not yet understood. Consequently, determining the QCD S-Matrix is a highly
non-trivial problem that is currently a long way from being solved and in
which, because the necessary analyticity properties should be present, 
S-Matrix Theory may yet have a role. 

In it's formal foundation during the early sixties, S-Matrix 
Theory was the name given to a broad set of principles and assumptions 
involving analyticity properties that
were postulated to be sufficient to dynamically determine the strong
interaction S-Matrix (Chew, 1962; Stapp, 1962;
Gunson, 1965; Eden et al., 1964). 
These principles were mostly extracted from quantum field
theory but were supposed to be able to stand alone. In fact, since in
isolation they appeared rather abstract and mathematical there was, from
the outset, considerable argument as to whether these principles constituted
a complete physical ``theory'' in any conventional sense. 

{\it ``... the possibility of analytically continuing a function into a 
certain region is a very mathematical notion, and to adopt it as a 
fundamental postulate rather than a derived theorem appears to us to be 
rather artificial.''} (Mandelstam, 1961)

{\it ``... when you find that the particles that are there in S-Matrix
Theory ... satisfy all these conditions, all you are doing is showing that
the S-Matrix is consistent with the world the way it is, ... you have not
necessarily explained that they are there.''} (Low, 1966)

\noindent Nevertheless, the early (pre-QCD) period during which S-Matrix
Theory was in it's ascendancy is a colorful, philosophical, and even
prophetic, part of the past that we will briefly review to provide
historical perspective. As a self-contained dynamical framework, S-Matrix
Theory probably reached it's peak with the formulation of dual resonance
models, which then evolved into the string theories that are the basis for
much of particle theory today. However, we will not attempt to bring such
theories under the umbrella of what we call S-Matrix Theory. 

Out of the initial formative period there also grew a smaller, but more
mathematical, school of ``Axiomatic S-Matrix Theory'' (Iagolnitzer, 
1976a,1978,1981,1993; Stapp, 1976a)  that
had less commitment to dynamical applications but rather concentrated on
establishing  analyticity properties directly from S-Matrix principles that
were clearly formulated independently of field theory. Eventually a 
classical correspondence principle for the S-Matrix, called macrocausality, 
was shown to lead directly to, and in fact to be equivalent to, local
analyticity properties. A number of important results were obtained using
macrocausality as a basic starting point that will surely carry over into
whatever the future holds for the subject. Although there have not been any
recent developments, we will outline the basic results
with emphasis on those that we anticipate will have a major role in the most
immediate future. 

While there is little current activity in formal 
four-dimensional S-Matrix Theory, in two space-time
dimensions there is a very active ``Exact S-Matrices'' field of research. 
This field is particularly interesting because, 
in theories that defy exact solution in the field theory
sense, the abstract S-Matrix principles have led to the discovery of
complete formulae for S-Matrices. Nevertheless, these results do not have any 
direct relation to the four-dimensional formalism with which this review is 
primarily concerned. Therefore, because of length limitations, 
we will not discuss these results in detail but rather refer the interested
reader to a number of good reviews that already exist in the
literature (Dorey, 1998; Mussardo, 1992; Zamolodchikov and Zamolodchikov, 
1979). 

A central theme which dominates the latter part of the article is 
that the asymptotic multi-regge region is where (at present) the general 
properties of the S-Matrix appear to be the most powerful. In this kinematic 
regime S-Matrix amplitudes
have a simple analytic structure that closely reflects the primitive
analyticity domains of local field theory (Epstein et al., 1976, Cahill and 
Stapp, 1975). The physical region
discontinuity structure needed to write multiple dispersion relations is
established within S-Matrix Theory (Stapp, 1976a,1976b) and the relationship
to the causality properties of field theory Green's functions is understood.
As a consequence multiparticle complex angular momentum theory, as proposed
and developed primarily by Gribov and collaborators (Gribov, 1962,1969; 
Gribov et al., 1965), can be put on a firm foundation (White, 1976,1991). 
Most importantly, the regge limit S-Matrix is controled by unitarity equations 
formulated directly in the angular momentum plane 
(Gribov et al., 1976; White, 1976, 1991). 
By design, these equations are satisfied by Reggeon Field Theory (Gribov, 
1967,1968; Abarbanel et al., 1974). The ``Critical Pomeron'', formulated 
using Reggeon Field Theory, provides the only
known non-trivial unitary high-energy S-Matrix (Migdal et al., 1974; Abarbanel 
and Bronzan, 1974). 

In considering future applications, we note first that since the S-Matrix is
surely the most important if not (in a broad sense) the only physical
quantity that needs to be calculated in a theory, any formalism
which helps in this purpose can be expected to retain some general
usefulness. There have been, and will continue to be, 
direct applications to spontaneously-broken non-abelian 
gauge theories, particularly in the regge region. Theories of this
kind are of interest both for the electroweak sector of the Standard Model
and for providing an infra-red regulated version of QCD. When the gauge
symmetry is broken completely by fundamental representation scalars the 
theory can be formulated gauge-invariantly (Banks and Rabinovici, 1979; 't 
Hooft, 1980).
There are no infra-red problems, all particles are massive, and both the
global analyticity properties of abstract field theory and pure S-Matrix
results apply to any finite order of the perturbative expansion. The most 
elaborate perturbative regge region calculations have been done in
such theories using, essentially, S-Matrix techniques (Fadin et al., 1977;
Fadin and Lipatov, 1996; Bartels, 1993) and this will 
surely continue into the future. Indeed, regge behavior is as definitive an
S-Matrix property of a perturbative massive non-abelian gauge theory
(Grisaru and Schnitzer, 1979) as is the much more widely appreciated
off-shell property of asymptotic freedom. The non-perturbative 
formulation of such theories has the serious problem that 
the self-interactions of the scalar fields involved are not
asymptotically free and so are uncontrollable in the
ultra-violet region. 

The current interest of most theorists is in going beyond the
Standard Model. Given the wide array of supersymmetric and grand unified
field theories, string theories, and most recently M-theory, that make up
the playbook for a particle theorist looking for interesting phenomena
outside the Standard Model, any immediate relevance for an abstract 
technical formalism that applies only to a (four-dimensional) S-Matrix seems
unlikely. If a better understanding of QCD is needed before the correct
extension of the Standard Model can be found then a role for S-Matrix Theory
in the search might be argued for. 

In general we anticipate that the main future applications for S-Matrix 
Theory will be to QCD, even though some sociological resistance may be 
encountered. Recent reviews of the discovery and development of
QCD have tended to portray the preceding S-Matrix era (and by implication
any succeeding S-Matrix formalism) as fruitless and outmoded following the
advent of QCD (Gross, 1999; 't Hooft, 1999). In some small part, 
this is probably a reaction to the unconcealed desire of some of the early,
most passsionate, S-Matrix advocates to proclaim the uselessness (if not the
death) of local field theory. 

{\it `` ... let me say at once that I believe the conventional association of 
fields with strongly interacting particles to be empty. ... no aspect of 
strong interactions has been clarified by the field concept.''} (Chew, 1962)

\noindent We should note that not all S-Matrix theorists adopted
this last point of view, even in the early days. 

{\it ``Some physicists have suggested retaining only the S-Matrix in our 
theory and discarding the remainder of the field theoretic framework .... 
One cannot then carry through the proofs of dispersion relations and 
analyticity properties, since the concepts on which they are based, ...
are essentially field theoretic ... it is not at all evident to us that all
physics is contained in the S-Matrix.''} (Mandelstam, 1961)

Nevertheless, upon it's discovery, QCD was seen as the triumphant return of
strong interaction field theory that should clearly vanquish all opponents
and all alternative formalisms (see for example De Rujala et al., 1975).
As we have already noted, the
crucial property of asymptotic freedom confounded the belief that a
consistent short-distance field theory interaction could not be found. Even
so, from the viewpoint of previously existing axiomatic formulations, where
it was always envisaged that there would be some direct relationship between
the fields and the asymptotic states, QCD is a very radical and
unconventional field theory.  The 
gauge-dependence of the fields is a deep property that is intricately
involved in making the short-distance part of the interaction finite but
that has to be absent in all physical quantities. All the problems of the
theory are at large distances. The S-Matrix of particles corresponding to
the fields does not exist, even in perturbation theory, because of infra-red
divergences. Conversely the physical particle states correspond to
complicated field configurations that may be only indirectly related to
local operators. 

Although it remains unproven, the uncontrollable infra-red behavior of the
QCD perturbation expansion is widely believed to be resolved by the
confinement of color charge, implying that the underlying fields are 
definitively unobservable. Almost invariably, though, the analyses and
arguments for confinement are made in the context of the euclidean path
integral on which modern (non-perturbative) field theory is based. 
Given the bad infra-red properties of perturbation theory, obtaining an
appropriate Minkowski space of physical states and a unitary S-Matrix is
an additional intricate demand\footnote{It is
only rarely acknowledged (see, for example, Weinberg, 1996) that the
non-perturbative euclidean formulation of a field theory contains no inbuilt
guarantee of unitarity in Minkowski space.} that surely is unlikely to be
satisfied for each and every non-abelian gauge theory, as current wisdom
would appear to imply. 

{\it ``We now know that 
there are an infinite number of 
consistent S-Matrices that satisfy all the sacred principles. 
One can take any non-abelian gauge theory, with any 
gauge group, and many sets of fermions ... ''} (Gross, 1999)

There is currently a wide body of knowledge on the non-perturbative
properties of field theories, most notably exploiting the lattice
formulation or, in the case of supersymmetric theories, duality properties
relating large and small momentum regions. In general, however, 
any possible particle S-Matrix 
remains far away from the domain of applicability of known non-perturbative 
results. Certainly, no S-Matrix has been directly
calculated, or even shown to exist outside of the perturbation expansion,
for any four-dimensional theory. Indeed the four-dimensional 
(infinite volume) euclidean
path integral also has yet to be proven to exist. 
At this point we would like to refer to a
comment of Feynman that was made nearly forty years ago but remains every bit
as true today (particularly if 
the roles of field theory and S-Matrix theory are interchanged, as we have 
taken the liberty of doing). 

{\it
``During all this time, no complete solution either of} the field equations
{\it or of } the S-Matrix {\it has really been produced. You sit there and
say: why isn't everybody doing} field theory; {\it another guy says: why
isn't anybody doing} S-Matrix theory? {\it The real problem is: WHY IS
NOBODY SOLVING ANYTHING? 

One of the reasons why you don't solve the problems is that you don't work 
hard enough. One of the reasons it is and has always been difficult to work 
hard on these problems is that nature keeps telling us it has the quality of 
being much more elaborate than we thought.''} (Feynman, 1961)

\noindent It could yet be that
the pathological properties of QCD as a field theory will require that
special S-Matrix based techniques be developed to directly construct the
physical hadron S-Matrix. In this case, the more balanced view of those who
remained agnostic in the heat of the idealogical battle between field theory
and S-Matrix theory may yet be seen as having long-term truth. The following
comment was made in the midst of the discovery of QCD. 

{\it ``The quarks inside a hadron seem to be very light, and almost free. Yet
they are not produced, even in very high energy reactions. This is an
unresolved mystery. Nevertheless, if we imagine that a resolution can be
found, we may then have a meeting of field theory and nuclear democracy: all
observable states will be bound states of quarks, and thus none will be
elementary; the dynamics of the quark fields will determine the hadron
spectrum, and the fields will make up the currents which carry the weak and
electromagnetic interactions. S-Matrix theory will describe the particles
from the outside, and field theory from the inside. There need be no
contradiction.''} (Low, 1974) 

\noindent In fact, without the parton model, which is partially an
embodiment of this last sentiment, very little of the hadron S-Matrix would
be calculable with present knowledge. Unfortunately, the parton model has a
solid basis only in the very limited kinematic regimes where leading twist
perturbation theory can be consistently formulated. 

Given the extensive regge region results referred to above,
an obvious question for the immediate future is
whether S-Matrix Theory can be used to obtain the regge limit of the QCD
S-Matrix. A-priori this problem is, of course, much simpler than determining
the full S-Matrix. It is also close to, although just beyond, the reach of
perturbation theory. As part of the argument for the futility of the
S-Matrix era it has been said that the regge region was unduly emphasized
and that today it is merely ``an interesting, unsolved and complicated
problem for QCD'' (Gross, 1999). In fact this is the kinematic regime where the
well-understood formalism of short-distance perturbative QCD comes closest
to a direct confrontation with unitarity and the infra-red spectrum. 

Using the language of reggeon diagrams, the regge region unitarity equations
can be used to organize and predict to all-orders  results
obtained from direct perturbative calculations within QCD
(Fadin et al., 1977; Bronzan and Sugar, 1978; White, 1993). It may then be
possible to use a Reggeon Field Theory phase-transition formalism (White, 
1991), with the U(1) anomaly providing a crucial ingredient (White, 1999),
to connect the reggeon diagrams of QCD to the Critical
Pomeron. Such manipulations could surely be described as ``state of the art
perturbation theory'' (the general title of this Section) in the sense that
S-Matrix Theory would be used to ultimately define the all-orders meaning of
the QCD perturbation series. 

Because of it's unique proximity to the perturbative domain, the high-energy
multi-regge region is probably the only kinematic regime where it might be
possible to construct the physical S-Matrix (almost) directly from the
perturbation expansion involving elementary fields. However, since all the basic
properties of the physical spectrum, including confinement and chiral
symmetry breaking, must be present in the regge region, an understanding of
the asymptotic S-Matrix could be a very important step towards obtaining the
full S-Matrix. Indeed, that a consistent unitary result be obtained is
likely to be a strong constraint on the formulation of QCD which may even
restrict the quark content of the theory. Building on this, at the end of
this article, we very briefly discuss the extent to which the abstract
properties that must be satisfied are so powerful that the strong
interaction S-Matrix and, perhaps, even the full S-Matrix containing the
electroweak interaction, might be determined by these properties.

\section{The Early Years.} 

\setcounter{equation}{0}

\subsection{Living Without Field Theory. }

S-Matrix Theory was born during a period when particle physicists were
intensely scrutinising and reformulating all principles and concepts.
The renormalizability of Quantum Electrodynamics had been established but
the procedure left many dissatisfied. Moreover, it was suspected that the
weak-coupling expansion did not define the theory at large coupling and 
that, more generally, in any local field theory the growth of the
interaction at short distances produces ``nullification'' (or triviality, in 
current language) in that a finite result is obtained only 
with a zero bare coupling (Landau and
Pomeranchuk, 1955). 
There was a collective sentiment that field theory either could not be used at 
all to describe the strong interaction or that it would have to be 
formulated in an abstract, non-perturbative, even non-lagrangian, framework 
to be applicable. Simultaneously, dispersion relations were increasingly
successful phenomenologically in relating strong
interaction scattering processses (Goldberger, 1961) and it became clear 
that the parameters of a field theory could as well be introduced via
dispersive calculations as in low-order diagrammatic calculations
(Mandelstam, 1958). In
addition the use of dispersion relations for S-Matrix amplitudes did
not involve the off-shell momentum regions of (low-order) Feynman diagrams
that produced the infinities needing infinite renormalization. 

To compare with experiment the S-matrix is all that is needed (Landau, 
1959). The question was, of course, what 
determines the S-Matrix and how is a general framework to calculate it to be
implemented? The central idea was that the strong interaction S-Matrix is
unique and self-consistent. Therefore if enough general properties are
specified it will be uniquely determined (Chew, 1962).
The initial properties chosen are all satisfied (formally) in the
perturbation expansion of a local field theory (of massive particles). 

\subsection{Initial Postulates}

If all the particles are massive then it can be self-consistently assumed 
that the interactions are short-range and that
the set of all free particle states forms a complete set. 
The S-Matrix is the sum total of scattering amplitudes
for any initial configuration of free particles, of any kind, to scatter
into any final configuration of free particles. 
The following were the first postulates made. 

\noindent {\bf [1] ~Lorentz Invariance}. This is straightforward.

\noindent {\bf [2] ~Maximal Analyticity}. The origin of global analyticity in
the causality of local field theory is described in sub-section 2.5 below.
Maximal analyticity, as a postulated principle, says that S-Matrix
amplitudes are analytic functions that have only the minimal singularity
structure consistent with postulates {\bf [3]} and {\bf [4]}.

\noindent {\bf [3] ~ Crossing}. Within the complex mass-shell, the momentum
of an incoming particle can be continued (or crossed) to that of an outgoing
particle. This postulate asserts that (provided the crossed scattering
process is physically possible) the corresponding amplitude is obtained by
the same continuation. Thus, one analytic S-Matrix amplitude describes many
different physical regions. 

\noindent {\bf [4] ~ Unitarity}. Obviously, unitarity must be 
satisfied in all physical regions. Maximal analyticity thus implies that
each amplitude has only the Landau singularities (described in the next 
Section) associated with
physical thresholds in each of the physical regions in which
it describes the scattering. In formulating the unitarity equations cluster
decomposition, as described in the next Section, is assumed to hold, 
although this property actually follows
from the macrocausality principle adopted in Axiomatic S-Matrix Theory. 

\subsection{Further Postulates}

Historically, the postulates of subsection 2.2 were first implemented
via dispersion relation calculations (Goldberger, 1961; Chew and Mandelstam, 
1960). For four-particle scattering
amplitudes Mandelstam's double dispersion relation
provided a
formalism that was consistent with, and at least as powerful as, one-loop
Feynman diagram perturbation theory (Mandelstam, 1958,1961). Nevertheless,
additional postulates were necessary to make practical progress. 

\noindent {\bf [5] ~  The Bootstrap}. This asserted that one could
self-consistently insert particle and resonance states in one chanel and
discover, via crossing and unitarity, particles and resonances in another
channel. 

\noindent {\bf [6]~ Maximal Analyticity of the Second Kind - Nuclear Democracy}.
This asserts that all particles lie on regge trajectories.
This not only eliminated the need for subtractions (as additional parameters) 
in dispersion relations 
but also implemented the philosophical concept that 
no particles are to be regarded as elementary. All 
particles are bound states of 
each other and essentially the same as resonances. 

With these last two postulates included quite elaborate consistency
calculations were performed (Chew et al., 1962). The approach floundered, in
part, because the more complicated analyticity properties of higher
multiparticle amplitudes do not allow any analogue of the Mandelstam
representation. As we noted in the Introduction, the absence of such a
representation is fundamental because it prevents an iterative 
perturbation expansion using discontinuities plus dispersion relations.

Both {\bf [5]} and {\bf [6]} were prophetic in defining essential
properties for present day theories. It can be shown that in 
a spontaneously-broken gauge theory all vector bosons and fermions lie on
Regge trajectories if, and only if, the gauge symmetry is non-abelian
(Grisaru and Schnitzer, 1979). The 
self-coupling of gauge fields is the essential interaction that is needed,
just as it is for asymptotic freedom. Therefore {\bf [6]} anticipated the
existence of non-abelian gauge theories. It is also interesting that 
reggeization of an elementary particle is the only known mechanism within 
field theory for generating an isolated Regge
pole trajectory as both hadron and pomeron 
trajectories apparently are. Bound state Regge poles are typically 
generated with accumulations points of trajectories.
Thus, the reggeization of quarks and gluons in QCD could 
play an important role in determining the pomeron and hadron spectrum.

The bootstrap postulate was eventually given stronger dynamical substance via 
the concept of duality. The 
contribution of resonances (at low and medium energy) should approximate (or 
be dual to) the contribution of regge poles associated with the resonances
in crossed channels (Dolen et al., 1968). This led first to the idea of a
narrow resonance approximation in which hadrons would lie on linear regge
trajectories (Mandelstam, 1967) and then to the formulation of dual
resonance models, which were soon reformulated as string 
theories (for a review see Mandelstam, 1976). Therefore {\bf [5]} anticipated
dual resonance models and, ultimately, the existence of string theories. 

The approximate linear regge trajectories 
observed experimentally were thus understood to imply that 
hadrons are extended objects which in massive, high angular momentum,
configurations are string-like. While this is understood qualitatively in QCD 
as due to the color flux tubes produced by the confinement of color, there 
is no established string approximation to the hadron S-Matrix. It remains the 
greatest irony of present-day particle theory that string theories are now 
thought to primarily describe physics at the Planck scale of the 
gravitational interaction. There is, as yet, 
no experimental evidence at all for this association. It is based entirely 
on arguments of theoretical consistency. In sharp contrast, 
there is overwhelming experimental evidence that the hadron S-Matrix has
string-like properties and yet we have only a very qualitative picture as to
how this relates to QCD and confinement. It remains plausible, therefore,
that the hadron (i.e. the QCD) S-Matrix will someday be directly 
calculated via some form of string theory.

\subsection{QCD and a Final Postulate}

It is fair to say that,
apart from the approximate validity of chiral symmetries, properties of the
hadron S-Matrix played 
very little role in the discovery of QCD (although the existence 
of a qualitative string picture certainly contributed to it's acceptance).
More important were the external currents that are required by the existence of 
electroweak interactions - even if the hadron S-Matrix 
could be self-consistently bootstrapped. As Feynman (1967) showed
the algebra of currents (and their amplitudes) proposed by 
Gell-Mann is exactly what has to be added to the S-Matrix (Gell-Mann 1962,
see Fritzsch and Gell-Mann, 1973 for a full review). The spectrum of
hadrons suggested the currents 
should contain fractionally-charged quark fields. While Gell-Mann's attempts
to formulate the strong interaction in terms of currents were ultimately
unsuccessful he, and many others, argued that the desired currents would be
obtained from a strong interaction vector ``gluon'' theory (Fritzsch and 
Gell-Mann, 1973). Finally, it was realized that (confined) SU(3) color would
simultaneously explain the existence of baryons, the magnitude of the
$e^+e^-$ total cross-section, and the $\pi^0 \to 2 \gamma$ decay (Bardeen et
al., 1973). In parallel (almost) the 
experimental discovery of scaling in deep-inelastic electron-proton
scattering at SLAC required the existence of a parton model that, with the
theoretical discovery of asymptotic freedom, QCD was able to provide
(Gross and Wilczek, 1973,1974; Politzer, 1974). 

Historically, therefore, it was properties of the additional electroweak
interactions that actually led to the discovery of the
field theory needed to describe the ``inside of hadrons''. The jet physics
that could, perhaps, have led to the discovery of asymptotic freedom and
short-distance scaling in purely hadronic interactions was developed only
later. If self-consistency properties of the strong interaction S-Matrix
determine it's uniqueness, as the founders of S-Matrix theory believed, it
was not evident in the formulation of QCD. 

There is one more 
additional postulate that was often included. We discuss it last because it
may be crucially related to asymptotic freedom. 

\noindent {\bf [7] ~ Maximal Strength of the Interaction - the Total 
Cross-Section does not Fall Asymptotically}.
If a vector interaction appears perturbatively at high
energies and large momentum transfers then (barring subtleties of limits)
the cross-section from such processes, and therefore the total 
cross-section, will not decrease asymptotically.
Thus the maximal strength
postulate may be the S-Matrix equivalent of assuming that an
asymptotically-free non-abelian gauge theory underlies the strong
interaction. At the end of this article we will suggest that QCD indeed may
be determined by this property of the hadron S-Matrix but, as the Feynman
quote in the Introduction suggested, for this we have to work much
harder ! 

\subsection{Analyticity in Field Theory}

While global analyticity may be a counter-intuitive property to assume in the 
abstract, it is well-known to be a natural consequence of causality
in a local field theory of massive particles
formulated via canonical quantization in Minkowski space. In a theory of 
this kind it is either assumed that the fields satisfying local 
commutation relations also create the particles or, equivalently, that 
fields defined via the particle states satisfy local commutation 
relations (Lehmann et al., 1957). (For gauge theories, this formalism applies 
directly only when the gauge symmetry is 
spontaneously-broken and then only within perturbation 
theory.) To illustrate how analyticity properties are derived and, in 
particular, because it will be interesting 
to relate the asymptotic analytic structure discussed in Section 4 
back to the basic analyticity domains of field theory, we briefly 
describe the relevant formalism in this subsection.

If $\phi(x)$ is a space-time
field operator which creates (or destroys) the particles of the theory from
(or into) the vacuum, (micro)causality implies that such fields commute
at space-like separations, i.e. 
$$
\left[\phi(x),\ \phi(y)\right]=0\qquad (x-y)^2<0
\auto\label{4.8}
$$
By using reduction formulae (that ``amputate'' external propagators), the 
momentum-space 
$S$-Matrix amplitudes of the theory can be obtained from ``Generalized Retarded
Functions'' (GRFs) that are the Fourier transforms of
``retarded'' Greens functions (Epstein et al., 1976). The simplest example
of a GRF (which gives only a propagator rather than an S-Matrix element) is 
$$
G(p)=\int d^4x ~e^{ip\cdot x}\langle 0|\theta(x_0)\phi(x)\phi(0)|0\rangle,
\auto\label{4.9}
$$
(where $\theta(y)=1$, $y>0$; $\theta(y)=0$, $y<0$). Provided that the retarded
Greens functions are well-defined distributions---that is they are polynomially
bounded---then their Fourier transforms have extensive analyticity domains
because of the convergence provided by the Fourier exponential factors of
the form $\exp\left[ip\cdot x\right]$. For example $G(p)$ is manifestly
analytic in the ``forward-tube''
$$
({\rm Im}\, p)^2>0\qquad {\rm Im}\, p_0>0.\auto\label{4.10}
$$
This is referred to as the ``primitive'' analyticity domain of $G(p)$ since 
it follows directly from microcausality without any further analytic 
completion.

A complete set of GRFs can be
defined from all field products in the theory (Epstein et al., 1976). Each
GRF is analytic in a particular 
tube or ``cone'' which is a generalization of (\ref{4.10}). 
For an $N$-point function a general tube is defined by
$$          
\left(\sum_\Delta {\rm Im}\, p_i\right)^2>0\qquad \sum_\Delta ~{\rm Im}\,
p_{i0}\raisebox{1.5mm}{\centerunder{$\;\scriptscriptstyle >\;$}{$\;
\scriptscriptstyle <\;$}}
0\quad \forall
\Delta, \auto\label{4.11}
$$
where $\Delta$ is any channel, that is any subset of the external momenta
$p_1\ldots p_N$, and (\ref{4.11}) must be satisfied with either the $>$
or the $<$ sign operative in the second term for all $\Delta$. By use of the
``Edge of the Wedge'' theorem it is straightforward to extend analyticity
within the tubes to ``partial tubes'' in which any of the $p_i$ are real and
spacelike (Epstein, 1965). If all of the $p_i$ are real and spacelike then
all the N-point GRF's coincide and can be identified as analytic
continuations of one analytic (off-shell) N-point amplitude. Again the
combination of all tubes and partial tubes is referred to as the
``primitive'' analyticity domain of the N-point amplitude. 

As preparation for the development in later Sections, we also 
discuss here the relationship between the
primitive domains of analyticity and the kinematic variables that describe
regge behavior. Consider the amputated four-point function 
that gives the on-shell S-Matrix and suppose 
that $p_i^2 = - m^2 < 0, ~i=1,..,4$. Lorentz invariance allows us to 
go to a frame where the most general real momentum configuration is
$$
p_1=(0,m\sin\zeta,0,m\cos\zeta),\qquad p_3=(0,-m\sin\zeta,0,m\cos\zeta),
\auto\label{4.12}
$$
$$
\eqalignno{
p_2&=\left(m\sin\zeta \sinh\beta,\,m\sin\zeta\cosh\beta,0,-m\cos\zeta\right)
&\num\label{4.13}\cr
p_4&=\left(-m\sin\zeta\sinh\beta,-m\sin\zeta\cosh\beta,0,-m\cos\zeta\right)
&\num\label{4.14}\cr}
$$
If we allow $z=cosh \beta$ to be complex so that
$$                  
\eqalignno{
2\,{\rm Im}\,z={\rm Im}\,\cosh\beta&={\rm Im}\left[\rho e^{i\delta}+
{1\over \rho}e^{-i\delta}\right]=\left[\rho-{1\over \rho}\right]\sin\delta
&\num\label{4.15}\cr
2\,{\rm Im}\,\sinh\beta&={\rm Im}\left[\rho e^{i\delta}+
{-1\over \rho}e^{-i\delta}\right]=\left[\rho+{1\over \rho}\right]\sin\delta
&\num\label{4.16}\cr}
$$
then $p_1$ and $p_3$ remain real and spacelike while
$$
{\rm Im}\,p_2=-{\rm Im}\,p_4={m\sin\zeta\over 2}\sin\delta
\left(\left[\rho+{1\over \rho}\right],\left[\rho-{1\over \rho}\right]
,0,0\right).\auto\label{4.17}
$$
$$
=>~~~~\left({\rm Im}\,p_2\right)^2=
\left({\rm Im}\,p_4\right)^2>0\quad 
      {\rm Im}\,p_{20}=-{\rm Im}\,p_{40}\sim {\rm Im}\,z.
\auto\label{4.18}
$$
That is the cut $z$-plane 
is contained in the analyticity domain given by the GRF (partial) tubes.
Also, since 
$$
{\rm Im}~ s = {\rm Im}~ (p_1 +p_2)^2 = -2m^2sin^2\zeta ~{\rm Im}~z
\auto\label{ims}
$$ 
the cut $z$-plane corresponds to the cut $s$-plane. If we analytically
continue to $p_i^2  > 0,~\forall ~i$ then $\theta = i \beta$ can be
identified with the center of mass scattering angle for $p_1$ and $p_3$. For
higher amplitudes multi-regge theory introduces generalizations of $z$ 
which will similarly describe the (partial-)tube analyticity domains for
spacelike masses. 

Since each GRF also gives an $S$-Matrix element as a boundary-value from within 
it's tube, it is natural that the global analyticity properties of the 
off-shell N-point amplitudes should to extend to the S-Matrix. Indeed if 
off-shell N-point amplitudes also share the analyticity property of
finite-order perturbation theory, then there are no singularities in the
external mass variables that would block the continuation on mass-shell. To
prove this within the framework of 
Axiomatic Field Theory (without appealing to perturbation theory) is a 
difficult analytic completion problem which has only been carried through 
in detail for the four-point, and to a lesser extent, the
five-point function. The resulting single-variable dispersion relation 
was one of the triumphs of
the non-perturbative abstract formulation of field theory (Epstein, 1965).

The asymptotic analyticity properties used in Section 4 are obtained 
by combining the field theory results described above with complimentary
results from the S-Matrix Theory of Section 3. The field
theory cut-plane analyticity in $z$ discussed above is
illustrated in Fig.~2.1(a). 
\begin{center}
\leavevmode
\epsfxsize=5in
\epsffile{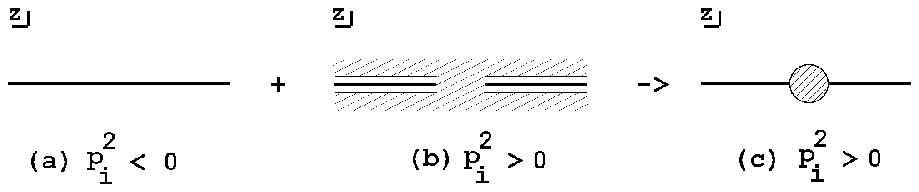}

Fig.~2.1 Analyticity Domains in the $z$-plane (a) in Field Theory (b) in 
S-Matrix Theory (c) Combining Both Formalisms.

\end{center}
In S-Matrix Theory, it is shown 
that when $p_i^2 >0$ there is a neighborhood 
of analyticity for the real $z$-axis of the kind illustrated in Fig.~2.1(b).
Complex Landau singularities are necessarily 
attached back to the normal threshold
branch cuts along the real axes. During the continuation from $p_i^2 < 0$ to 
$p_i^2 > 0$, such singularities can only 
emerge by some finite amount from the central part of the real axis.
Hence there must be a domain of analyticity as illustrated in Fig.~2.1(c)
which is the cut-plane minus some finite central circle where ``anomalous
threshold'' Landau singularities are (possibly) located. 
This argument generalizes
to multiparticle amplitudes when analyticity domains are discussed in terms 
of generalized $z$ variables.

\section{Axiomatic S-Matrix Theory. }

\setcounter{equation}{0}

In it's purest form S-Matrix Theory is 
devoted to establishing those analyticity properties of the S-Matrix that 
can be based on physical principles clearly 
formulated separately from field theory.
Results from this program will be the subject of this Section. The initial
hope was that such properties would be fed back into the dynamical program.
However, it soon became apparent that the subject was sufficiently
complicated that if the process of extracting and generalising results from
Feynman diagrams was to be abandoned, as a matter of principle, only rather
limited results could be obtained with any degree of rigor and generality. 

\subsection{Unitarity, Bubble Diagrams and Landau Diagrams }

The ``diagrammatic'' framework of S-Matrix theory is provided by the 
unitarity equations. A heuristic way to develop a
diagrammatic expansion that, superficially at least, is connected to the
Feynman diagram expansion is as follows. First write the $S$-Matrix
as $S=1+R^+$ and its Hermitian conjugate as $S^+=1-R^-$. 
The unitarity equation $SS^+=1$ can be written formally as 
$$
R^+~=~R^- ~[~1-R^-~]^{-1}~~=~\sum~(R^-)^n 
\auto\label{une}
$$
For simplicity we assume there is only one kind of
physical particle, a scalar with mass $m$. 
If we make the cluster decomposition $S(p_1,...,p_m;p_{m+1},...,p_n)~~= $
\newline \parbox{6in}{
\begin{center}
\leavevmode
\epsfxsize=5in
\epsffile{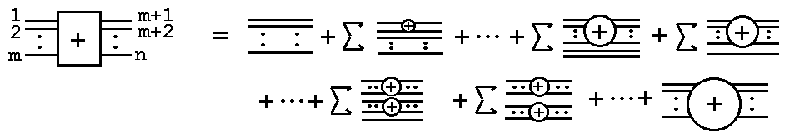}
\end{center}
}\parbox{0.5in}{$$~\auto$$}
\newline where a bubble represents a connected amplitude, together with a
momentum conservation $\delta$-function, then (\ref{une}) gives,
in bubble diagram notation,
\newline \parbox{6in}{\begin{center}
\leavevmode
\epsfxsize=5.5in
\epsffile{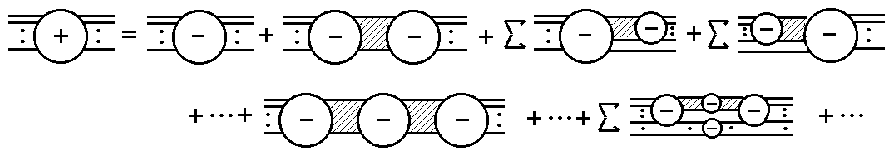}
\end{center}
}
\parbox{0.5in}{$$ ~ \auto\label{bbs} $$
}
\newline where the phase-space integration is a sum over all particle
numbers $N$ of intermediate lines, together with an integration over the
on-shell momenta involved.
\newline \parbox{1in}{$~$}\parbox{0.8in}{
\begin{center}
\leavevmode
\epsfxsize=0.6in
\epsffile{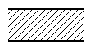}
\end{center}
}
\parbox{3in}{
$$
=~~~~\sum_{i=1}^N~~\int~~\prod_{i=1}^N~ {d^4 p_i \over (2\pi)^3}~ \delta^+(p_i^2 
-m^2)~\Theta
$$}
\parbox{1.7in}{$$~
\auto\label{psp}
$$
}
\newline where $\Theta$ is the inverse of the symmetry number of the 
state. 

Only ``$-$ bubbles'' appear in the expansion (\ref{bbs}). 
A general bubble diagram consists of signed bubbles connected by 
lines (directed from left to right). Each bubble diagram represents an integral
of the product of the functions corresponding to the bubbles within the 
diagram. The integration is over the on mass-shell momenta of each of the 
internal lines. Each bubble diagram function contains, therefore, an overall
momentum conservation $\delta$-function. 
At any finite energy each of the infinite series of bubble diagrams in
(\ref{bbs}) can be 
rearranged into a finite series involving both $-$ and $+$ bubbles by using
the unitarity equation for sub-sets of diagrams. The usual form of the
unitarity equation is then obtained, e.g below the four-particle threshold 
\newline \parbox{6in}{\begin{center}
\leavevmode
\epsfxsize=5in
\epsffile{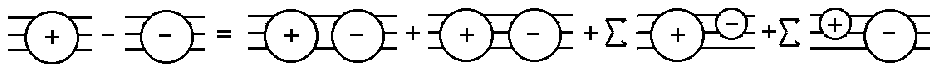}
\end{center}
}
\parbox{0.5in}{$$ ~ \auto\label{bbs1} $$}

The advantage of the series (\ref{bbs}) is that it displays the necessary
singularities of an amplitude explicitly. As the energy increases a new series 
of terms appears in the expansion whenever the threshold for a new 
scattering process is crossed. Scattering processes are described by a 
Landau diagram, examples of which are given in Fig.~3.1.
\begin{center}
\leavevmode
\epsfxsize=3.5in
\epsffile{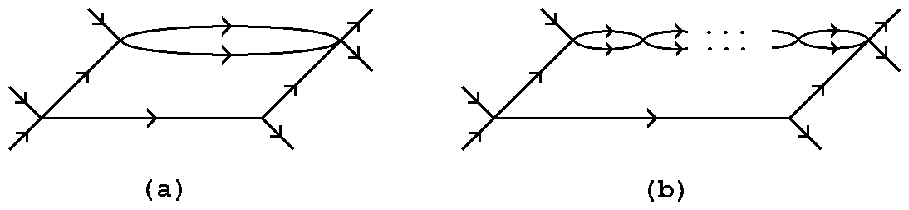}

Fig.~3.1 Landau Diagrams with (a) No Iterated States (b) Iterated States.
\end{center}
A Landau Diagram is composed of lines and vertices. Each line is
directed from left to right and carries a four momentum $p_i$ with
$p_i^2=m^2$. There is momentum conservation at each of the vertices. 
A Landau diagram ``fits into'' 
a bubble diagram if and only if the diagram can be constructed by fitting
into each bubble either a point vertex or a connected
(sub) Landau diagram. The Landau diagrams that represent threshold processes
in (\ref{bbs}) are obtained by replacing all the $-$ bubbles by point
vertices. 

\subsection{The Landau Equations, the ``$+~ \alpha$'' Condition, 
and Unitarity. } 

If a Landau diagram, such as that of Fig.~3.1(a), describes a space-time 
scattering process then for each internal loop $L_k$ of the diagram the
internal momenta $p_i$ (directed around the loop) must satisfy the condition
$$
\sum_{i~ \epsilon~ L_k}~~\alpha_i~p_i~~=~0
\auto\label{le}
$$
$\alpha_i \geq 0$ is required for the scattering to be physical. The Landau 
equations for a given Landau diagram are the set of conditions (\ref{le}) 
for all internal loops of the diagram. The set of external momenta 
that satisfy the complete set of equations provide the Landau singularity 
(surface) corresponding to the diagram(Landau, 1959). The set of all physical
region ($+ \alpha$) Landau diagram thresholds is the minimal set of
singularities that an analytic scattering amplitude must have. 

For a higher-order scattering process to be possible, all the sub-scattering 
processes must obviouly be above threshold. This leads to a heirachical 
property for $+\alpha$ Landau surfaces whereby surfaces of increasing 
complexity emerge from those of less complexity. In particular, if a 
complex Landau surface exists, it must be connected back (in general via
lower-order surfaces) to the real normal thresholds. Hence if a 
neighborhood of analyticity close to the physical region is proven to exist
as in Fig.~2(b), then the surfaces can only emerge from the low-energy 
ends of the normal thresholds, as allowed for in Fig.~2(c).

Note that if the external momenta are such that the scattering process of
Fig.~3.1(a) is possible, then clearly any process with an iterated
interaction of the form of Fig.~3.1(b) is also possible. The Landau
equations for the iterated loops simply imply that the two momenta involved
are parallel and no additional constraint is placed on the external momenta.
Hence all diagrams of this form will have the same threshold and will give
the same Landau surface. The nature of the singularity at the threshold
will nevertheless be different for each diagram. To obtain the complete
discontinuity around the surface it is necessary to include together all
iterated scattering processes. 

The Landau singularities were originally discovered in Feynman diagrams
(Landau, 1959). The purpose of S-Matrix Theory is, however, to derive the
analytic 
structure of amplitudes and, in particular, discontinuity formulae directly
from unitarity. Unfortunately there are two simplifications of Feynman
integrals that are not necessarily present in the bubble diagram integrals
appearing in unitarity equations. Because of the $+i\epsilon$ prescription
for propagators, Feynman integrals can be analytically continued around a
Landau singularity. Also, because the vertices are point interactions, 
singularities are generated only by propagators. As a consequence Feynman
diagram integrals are clearly singular only on the Landau surfaces obtained from
$+\alpha$ Landau equations. In S-matrix Theory there is, a-priori, no
$i\epsilon$ prescription. In addition, $-$ bubbles have the complex
conjugate singularity structure of the $+$ bubbles. Consequently, when 
the singularities of bubble diagram amplitudes within a bubble diagram are 
allowed for, singularities are generated on ``mixed-$\alpha$'' solutions of the 
Landau equations in which negative $\alpha$'s are assigned to those lines of 
the Landau diagram that are produced within a $-$ bubble. Mixed-$\alpha$
solutions of iterated interaction diagrams, such as that of Fig.~3.1(a), are 
particularly troublesome.

It is thought to be self-consistent to assume both the absence of physical
region mixed-$\alpha$ singularities and the $+i\epsilon$ 
prescription. In some of the early formulations of
S-Matrix Theory, the $+i\epsilon$ prescription for analytic continuation was
adopted as an additional postulate. At the same time there were several
attempts to develop an S-Matrix concept of causality (for a complete set of
references see Chandler and Stapp, 1969). 
These efforts culminated in the
formulation of an S-Matrix classical correspondence principle called {\it
macrocausality} that has as a consequence local analyticity properties that
include both the $+i\epsilon$ prescription and the absence of
mixed-$\alpha$ Landau singularities. 

\subsection{Macrocausality and Essential Support}

The macrocausality principle says that all interactions between particles
fall-off exponentially under space-time dilations unless the interaction can
be transmitted by the exchange of stable particles. It leads
to, or alternatively is equivalent to, the existence of (infinitesimal)
domains of analyticity for $S$-Matrix elements in the immediate neighborhood
of physical regions (Iagolnitzer and Stapp, 1969). Macrocausality can
therefore replace the microcausality of local field theory as a basis for
local (but not global) analyticity properties. While it is thought to be 
consistent with the microcausality property of field theory, a direct
relationship has not been established. It is believed that unitarity of the
$S$-Matrix has to be combined with microcausality to derive macrocausality.
This is effectively the purpose of the ``non-linear program'' of field
theory in which ``asymptotic completeness'' is added as an additional
axiom 
to allow the mixing of unitarity properties with the
existence of off-shell analyticity domains (Bros and Lassalle, 1976; 
Iagolnitzer, 1993). 

While the essential concept of macrocausality is straightforward, an exact
definition is more elusive and so we will not give one here. To formulate a
precise principle requires the introduction of concepts 
that are a-priori outside of momentum space S-Matrix Theory
(Iagolnitzer and Stapp, 1969; Iagolnitzer, 1976a,1978,1981). Firstly the
discussion of a particle's position in space-time requires the introduction
of appropriately localized wave-functions and there is room for variation in
this. Secondly, the characterization of the exponential decrease of the
probability for scattering in non-causal configurations allows some
variation. As a consequence differing, but essentially equivalent, proofs of
the $+i\epsilon$ prescription and the presence of only $+\alpha$ Landau
singularities can be given (Iagolnitzer and Stapp, 1969; Iagolnitzer, 
1976a,1978,1981). 

In general, when S-Matrix Theory is developed from macrocausality, the
notion of the essential support of a generalized
fourier transform plays a central role
(Iagolnitzer, 1976b,1978,1981). The generalized transform of a
``momentum space'' function $f(p)$, where $p=(p_1,p_2,...,p_n)$ is defined
as 
$$
F(x,p;\gamma)~=~\int ~e^{-ip'.x-\gamma|x||p'-p|^2}~dp'
\auto\label{es}
$$
The essential support of $f$ at $p$ is the set of ``singular'' directions 
along which $F(x,p;\gamma)$ does not decay exponentially in $x$-space. Some 
continuity properties are also required. The essential support of $f$ can 
also be viewed as the cone with apex at the origin in x-space formed by the 
singular directions. A relationship between local analyticity properties in 
momentum space and essential support properties in co-ordinate space 
holds that parallels the relationship between global
momentum space analyticity and conventional support properties in
co-ordinate space (Iagolnitzer, 1976b,1978,1981). 

\subsection{The Structure Theorem}

The basic result needed for obtaining discontinuity formulae from 
unitarity is provided by the Structure Theorem. The essence of this theorem 
is that if all the singularities of the sub-bubble functions contained
in a bubble diagram have only Landau singularities in the physical
region, then so also does the full diagram function. Moreover, the Landau
singularities of the full diagram are just those whose Landau diagram fits
into the the full bubble diagram. The theorem is most easily 
proved using essential support theory
(Iagolnitzer, 1976a,1978,1981). It then
states (loosely) that the essential support of a bubble diagram function can
be computed from the essential supports of the sub-bubbles. 

There are exceptions to the structure theorem for special momentum
configurations which, in principle, weaken the generality of the 
discontinuity formulae in the next two sub-sections. These exceptions allow 
very singular contributions from mixed-$\alpha$ Landau singularities, 
particularly those arising from iterated interactions such as appear in 
Fig.~3.1(b). As a
result, the cancelation of mixed-$\alpha$ singularities within the unitarity
equation, which is required by macrocausality for physical amplitudes, is
not proven. If the following discontinuity formulae 
are first derived assuming general mixed-$\alpha$ cancelations, it is 
possible to show, a-posteori, that the needed cancelations do take place.
However, another solution in which such cancelations are absent can not be
ruled out - even though there is no evidence, diagrammatic or otherwise, for
it. At present such (purely mathematical) road blocks to uniqueness have to
be presented as qualifications for the results of the following sub-sections. 

In the framework of essential support theory the assumption of
mixed-$\alpha$ cancelation becomes a stronger and clearer ``separation of
singularities'' assumption (Iagolnitzer, 1976a,1978,1981). This assumption also
leads to 
adaptations of the discontinuity formulae in situations where usual
discontinuity formulae can not be formulated. In general the proofs of
discontinuity formulae using essential support theory are much more
satisfactory than the elementary manipulations of bubble diagram expansions
that we use below which, however, give the results
in a direct and intuitive manner. 

\subsection{Local Discontinuity Formulae}

We first discuss the local discontinuity around a general $+\alpha$
Landau surface that appears as a threshold in the bubble diagram
expansion (\ref{bbs}). The structure theorem implies that the discontiunity
is given directly by the new terms that appear. To obtain a compact
discontinuity formula we rearrange each of the infinite
series due to iterated interactions using unitarity equations. 

As an example(Stapp, 1976a) which illustrates the general case, we consider the
Landau diagram of Fig.~3.2(a). 
The new terms in the expansion (\ref{bbs})
are all the diagrams having the general form shown in Fig.~3.2(b). 
After using unitarity equations we obtain the discontinuity as the
bubble diagram illustrated in Fig.~3.3. 
The letters $a$, $b$ and $c$ label sets of particles, which could obviously 
be different to those we have shown. The corresponding thin lines cut sets 
of internal lines that correspond to 
these sets of particles. 
\newline \parbox{2.7in}{\begin{center} 
\leavevmode 
\epsfxsize=2.2in
\epsffile{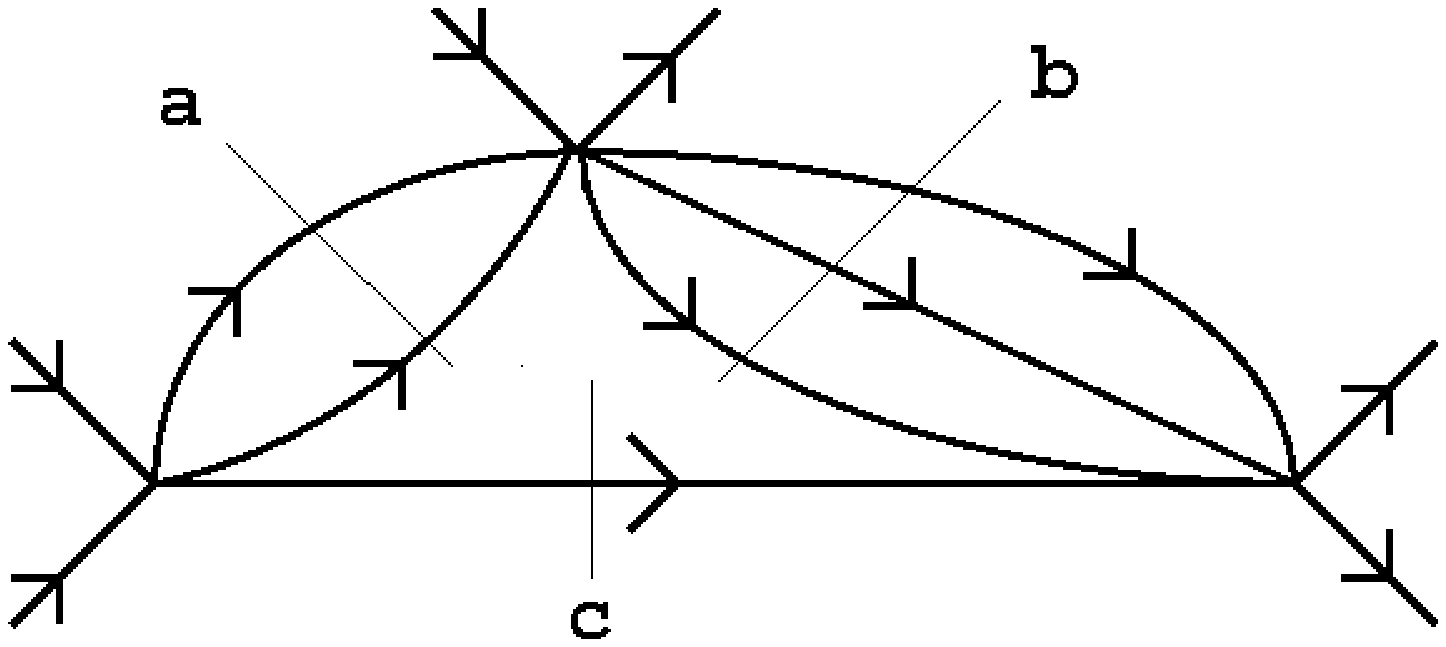}

(a)
\end{center}
}
\parbox{3.8in}{
\begin{center}
\leavevmode
\epsfxsize=3in
\epsffile{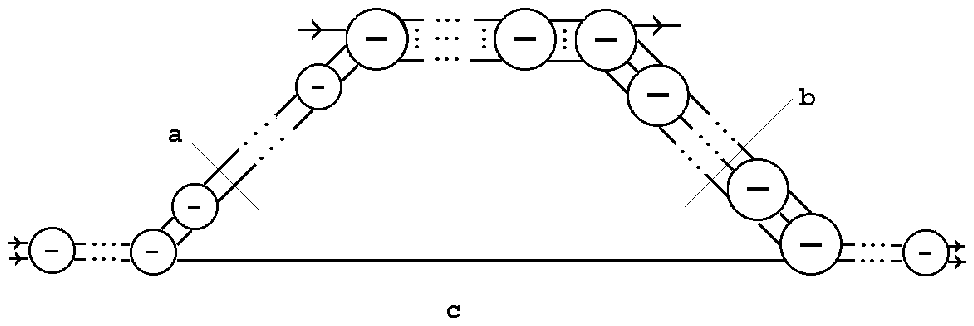}

(b)
\end{center}
}
\newline \centerline{Fig.~3.2 (a) A Landau Diagram (b) The Threshold Bubble 
Diagrams.}
\begin{center}
\leavevmode
\epsfxsize=2.5in
\epsffile{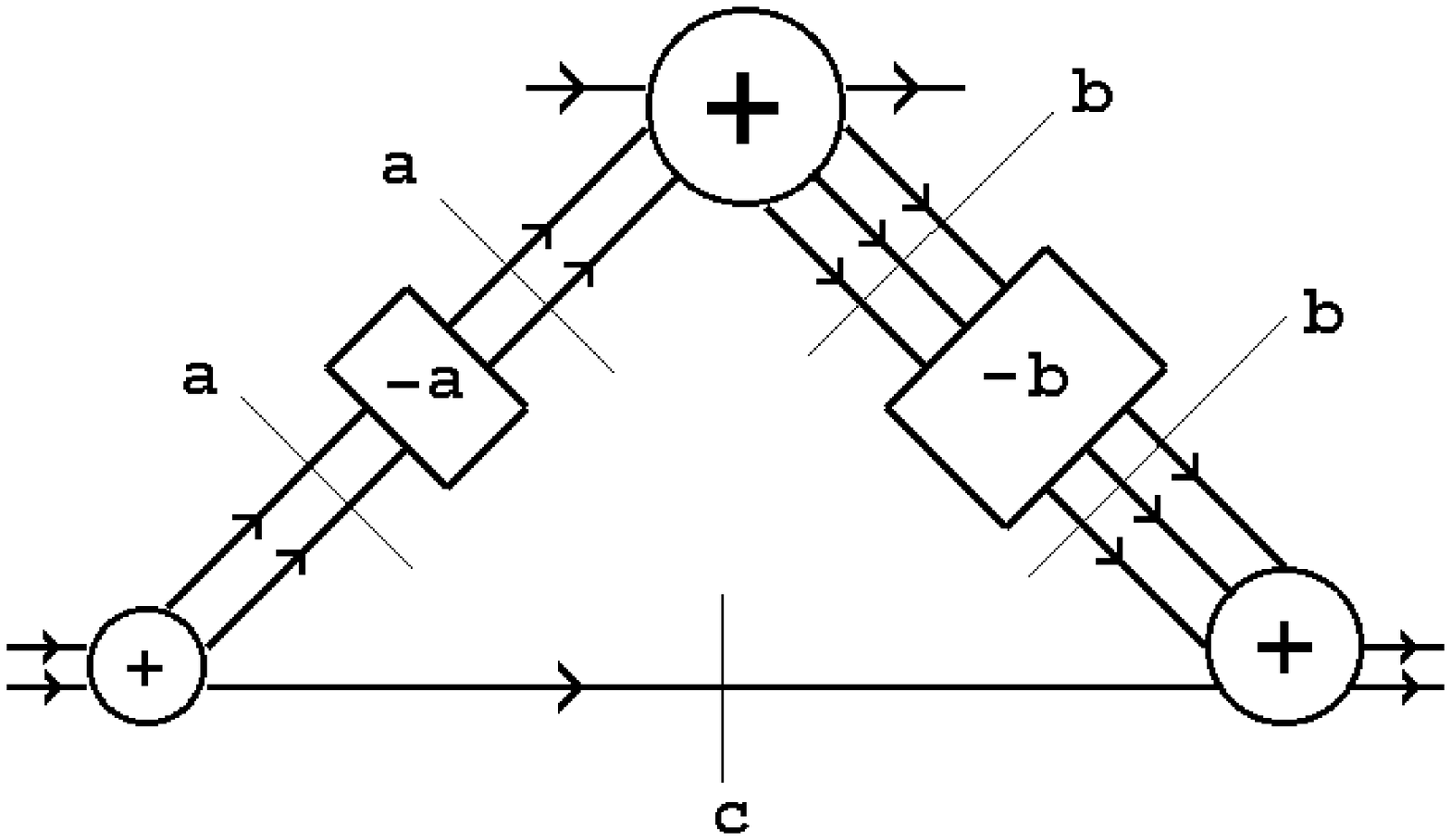}

Fig.~3.3 The Discontinuity for Fig.~3.2.
\end{center}
The $-a$ and $-b$ boxes 
represent the functions 
$S^{-1}_a$ and $S^{-1}_b$ defined by 
$$
S_i~S^{-1}_i~=~I_i      ~, ~~~ i~=~a,b
\auto\label{res}
$$
where $S_i$ and $I_i$ are restrictions of the S-Matrix $S$ and the unit 
operator $I$ to the space corresponding to the set of particles $i$.
 
\subsection{Good and Bad Functions and the Steinmann Relations} 

In addition to local discontinuities of the kind discussed in the last 
subsection, we can also discuss the total discontinuity in a particular
invariant and similarly multiple discontinuities with repect to several 
invariants. A-priori, the amplitude below all cuts in an invariant can be
defined by simply removing all bubble diagrams in the expansion (\ref{bbs})
that have an intermediate state in that invariant. In the same 
way amplitudes above the cuts in any set of invariants and below in any
further set can (naively at least) be defined. In addition to this algebraic
procedure there is another equally valid procedure. The bubble expansion
(\ref{bbs}) written with $+$ bubbles instead of $-$ bubbles holds also for
$S^+$, which should correspond to the amplitude evaluated below all
invariant cuts. Starting from this last expansion amplitudes above specific
invariant cuts should be obtained by removing terms from the expansion. The
the functions obtained by these two procedures may not coincide.

An additional complication arises from the lack of independence of
invariant variables. A particular problem is that it is possible to 
divide the invariants into two sets such that there is no mass-shell point 
that lies simultaneously in the upper half-plane of one set and in the lower
half-plane of the other set. For example, if three invariants $s_a$, $s_b$
and $s_c$ satisfy $s_a+s_b= s_c+ Cm^2$, where $C$ is a real constant, then
it is not possible to have Im$s_a$, Im$s_b > 0$ and Im$ s_c <0$. Such a
combination of imaginary parts is referred to as an ``inaccessable
boundary''. In principle, an (artificial) infinitesimal parameter can be added 
to the amplitude such that the cuts separate as in Fig.~3.4. 
{\begin{center}
\leavevmode
\epsfxsize=4in
\epsffile{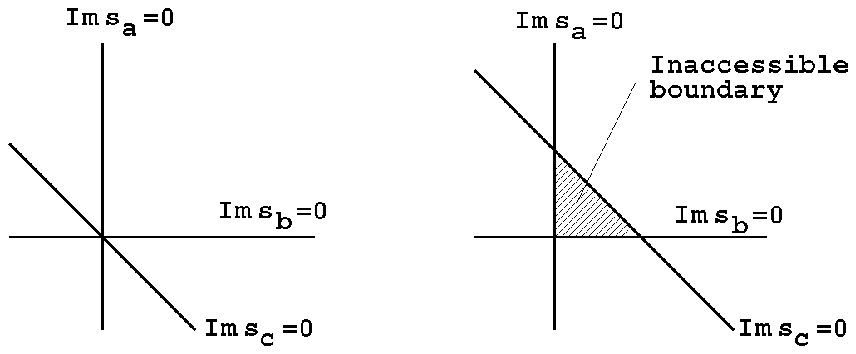}

Fig.~3.4 Separating Cuts to Expose an Inaccessible Boundary.
\end{center}
We can then define the amplitude in any way we 
choose in the 
inaccessible boundary and, as in the next Section, write a multivariable
dispersion relation, before returning the infinitesimal parameter to
zero.

With this last manipulation, amplitudes above and below all combinations of
cuts can be defined by the two algebraic 
procedures described. For some amplitudes, called ``good
functions'', the two possibilities coincide. The good 
functions are thus uniquely determined and, in fact, share the good analyticity 
properties of the physical sheet amplitude. (It is anticipated that they are
related, by some path of analytic continuation, to this amplitude.) Amongst
the good functions are all the amplitudes obtained from the most general set
of field theory ``Generalized Retarded Functions''
discussed in Section 2 (Cahill and Stapp, 1975). 
The four-point amplitude below the normal thresholds cut 
is both a good function and a GRF.
Therefore, it's good analyticity properties immediately translate into the 
analyticity domain illustrated in Fig.~2.1(b). A similar domain holds for a 
cut in a multiparticle amplitude whenever the amplitudes on both sides of 
the cut are good functions.

The ``bad functions'' are those for which the two
algebraic definitions do not 
coincide. As an example, for a $3-3$ amplitude the 
bad functions are from boundary-values (or the conjugates) 
in the combination of invariant cuts of the kind illustrated in Fig.~3.5 
\begin{center}
\leavevmode
\epsfxsize=1.6in
\epsffile{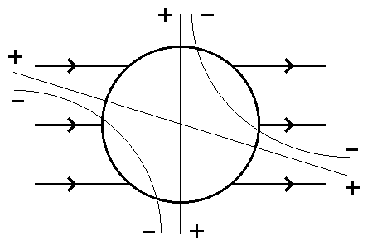}

Fig.~3.5 Bad Boundary-Values
\end{center}
- the 
thin lines indicate the invariant channels and the 
$\pm$ signs indicate the boundary-values. 
Not only are the bad functions not uniquely defined. Either definition gives 
very bad analyticity properties. There are complex cuts extruding from the 
real region and no path linking the amplitude on either side. There is,
however, a unique definition that is not equal to either algebraic
definition. The bad functions can be defined so that the Steinmann relations
are satisfied (Stapp, 1976a: Cahill and Stapp, 1975; Coster and Stapp, 1975).
This does not lead to better analytic
properties but leads to important properties for the asymptotic
dispersion relations of the next Section, as we discuss further below. 

The original Steinmann relations are satisfied by the GRFs of field theory 
and are direct consequences of microcausality. In the S-Matrix context the 
Steinmann relations are said to be satisfied if there are no double
discontinuities in overlapping channels. If the bad functions are
defined by simply omitting (from either bubble diagram expansion) those
diagrams that contain the Landau diagrams giving the undesired double
discontinuities, the Steinmann relations are satisfied. The resulting
functions still have bad analytic properties and are certainly not
related to physical sheet amplitudes by any path of analytic continuation,
but the simple discontinuity formulae of the next subsection are obtained.
A central point for
the next Section is that the bad boundary values become inaccessible in
multi-regge limits and so the Steinmann relations can be legitimately
imposed in writing asymptotic dispersion relations. The validity of
multi-regge theory is deeply tied to the Steinmann relations and their
validity in turn reflects the direct relationship between multi-regge region
analyticity and the primitive domains of field theory. 

\subsection{Global Discontinuity Formulae}

With the Steinmann relation definition of bad functions multiple discontinuity 
formulae are as would be obtained if all higher-order Landau singularities
were absent and there were only normal threshold branch cuts. To give some
simple examples we first introduce the additional diagrammatic notation 
of Fig.~3.6. 
\newline \parbox{3.1in}{ 
\begin{center}
\leavevmode
\epsfxsize=2.7in
\epsffile{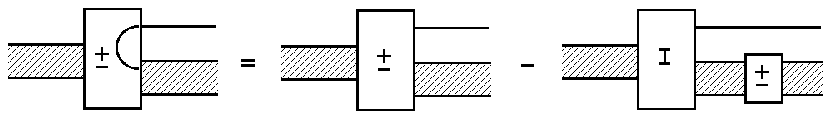}
\end{center}
}
\parbox{0.2in}{\begin{center} , \end{center}
}
\parbox{3.1in}{
\begin{center}
\leavevmode
\epsfxsize=2.7in
\epsffile{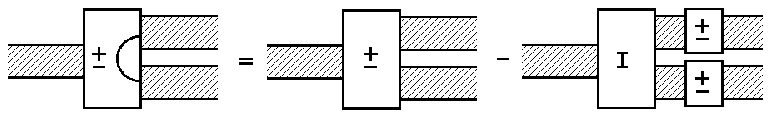}
\end{center}
}
\centerline{Fig.~3.6 Diagrammatic Notation}

\noindent Using this notation, together with the corresponding notation
involving initial state particles, the various triple discontinuities of a
six-point amplitude can be represented as in Fig.~3.7. 
The channels in which 
the discontinuities are taken are those in which the hatched lines appear. 
These formulae are
anticipated to generalize to the (N-3)-fold multiple discontinuities needed
for the asymptotic dispersion relations of the N-point amplitude discussed
in the next Section. 
\begin{center}
\leavevmode
\epsfxsize=4in
\epsffile{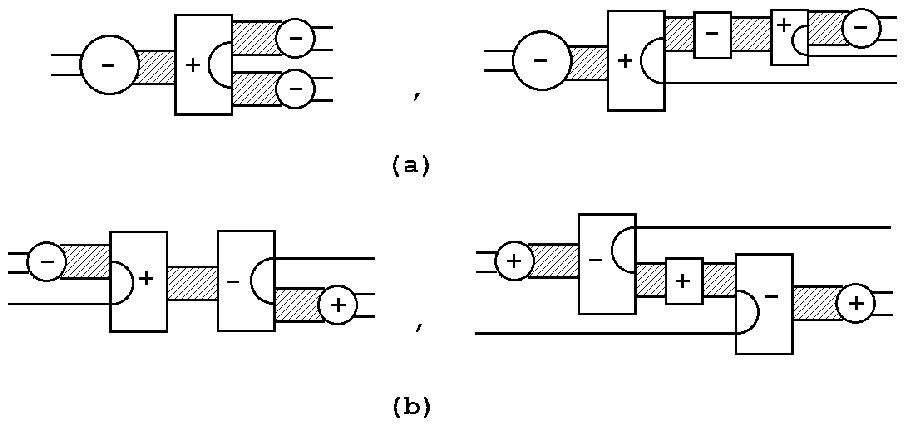}

Fig.~3.7 Triple Discontinuities for (a) 2-4 Scattering (b) 3-3 Scattering.
\end{center}

\subsection{CPT, Hermitian Analyticity, etc.}

We have not discussed at all the technique of using pole factorization 
to embed a low-order amplitude in a higher amplitude and thus use analyticity
properties of the higher amplitude to deduce properties of the low-order 
amplitude. Early results obtained this way were the CPT Theorem and hermitian
analyticity (Stapp, 1962; Olive, 1964). Paths of analytic continuation
relating good functions can also be discussed this way. 

\subsection{Holonomy}

This is a very interesting mathematical subject that we will not attempt to 
review (Iagolnitzer, 1978,1981,1993). It began with the suggestion
(demonstrated in some cases) that the discontinuity formulae satisfied by
$S$-Matrix elements could be regarded as infinite order ``pseudo-differential''
equations. It then appeared 
that the maximal analyticity assumption of S-Matrix Theory might be embodied
in Sato's conjecture that the $S$-Matrix is a holonomic
microfunction---that is it is a solution of a maximally over-determined
system of pseudo-differential equations (Sato, 1975). Presently this
conjecture seems to have been disproven, at least in it's
simplest form (Iagolnitzer, 1993). 

\section{Asymptotic S-Matrix Theory.}

\setcounter{equation}{0}

\subsection{The Elastic Scattering Asymptotic Dispersion Relation.}

Given the analyticity domain illustrated in Fig.~2.1(c), Cauchy's theorem 
implies that an 
elastic scattering amplitude $A(s,t)$ 
satisfies the single-variable dispersion relation 
$$
A(s,t)={1\over 2\pi i} \int_{I_R} {ds^\prime\Delta(s^\prime,t)\over
(s^\prime-2)} + {1\over 2\pi i} \int_{I_L} {ds^\prime \Delta(s^\prime,t)\over
(s^\prime-s)} + {1\over 2\pi i} \int_{\scriptstyle |s^\prime|=R\atop
\,\,\scriptstyle |s^\prime|=s_0} ds^\prime {A(s^\prime,t)\over (s^\prime-s)},
\auto\label{4.2}
$$
where $I_R$ and $I_L$ are the right and left cuts 
respectively. If we are interested only in the leading 
Regge behavior the third term can be ignored, provided the first two terms are 
evaluated appropriately. Suppose that 
$$
A(s,t)\,\centerunder{$\large\sim$}{$\scriptstyle|s|\to\infty$}\,\beta_+(t)
s^{\alpha(t)}+
\beta_-(t)(-s)^{\alpha(t)},\auto\label{4.4}
$$
(There could also be additional $\ln s$ dependence.) 
For ${\rm Re}\,\alpha(t) < -1$ we can use 
$$                                                          
{1\over 2\pi i}\int_{|s^\prime|=R}{ds^\prime A(s^\prime,t)\over (s^\prime-s)}
\,~\centerunder{$\large\sim$}{\raisebox{-1mm}{$\scriptstyle R\to\infty$}} 
\,~R^{\alpha(t)}\to
0\qquad {\rm Re}\,\alpha(t)<-1.
\auto\label{4.6}
$$
If we take $R\to\infty$ and then analytically continue 
to ${\rm Re}\,\alpha(t)>-1$, since

$$
{1\over 2\pi i}\int_{|s^\prime|=S_0}ds^\prime {A(s^\prime,t)\over (s^\prime-s)}
~~\,\centerunder{$\large \sim$}{$\scriptstyle|s|\to\infty$}\,~~O\left({1\over
s}\right),\auto\label{4.5}
$$
we can write
$$
A(s,t)={1\over 2\pi
i}\left(\int_{I_R= (s_0,\infty)}+\int_{I_L = (-\infty,-s_0)}\right)
{ds^\prime\Delta(s^\prime,t)\over
(s^\prime-s)}+A_0,\auto\label{4.7}
$$
where the $I_R$ and $I_L$ integrals are defined by analytic continuation and
$A_0$ gives sub-dominant asymptotic behavior. 
(\ref{4.7}) is the simplest example of an ``asymptotic
dispersion relation''. It differs from the exact dispersion relation
only in that details that are irrelevant in the regge limit are omitted. 

\subsection{Multiparticle Kinematics and Analyticity Domains.}

There is no simple generalization of (\ref{4.2}) to multiparticle 
amplitudes, in 
part because of the increased complexity of the singularity 
structure due to the large number of invariants. 
However, there is a generalization of (\ref{4.7})
(Stapp, 1976b; Stapp and White, 1982; White, 1991). 

The first step is the introduction of angular
variables for an N-point amplitude.  
\newline \parbox{3.1in}{Each of the many possibilities 
corresponds to a distinct tree diagram with three-point vertices - a ``Toller 
Diagram''.  As illustrated in Fig.~4.1 for the 6-point amplitude, 
internal momenta are introduced by imposing momentum conservation.
At each vertex, 
three Lorentz frames are defined 
in which each of the entering $Q_i$ take a standard form. 
Writing $t_i = Q^2_i$ we then have }
\parbox{3.4in}{\begin{center}
\leavevmode
\epsfxsize=2.6in
\epsffile{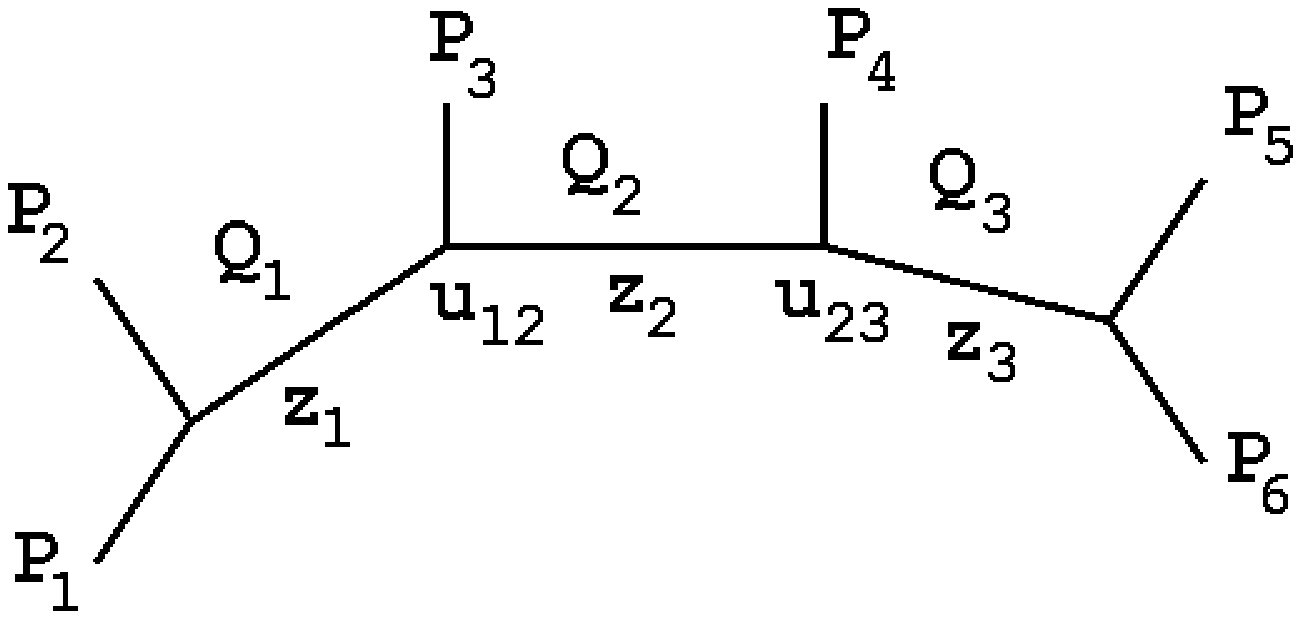}

Fig.~4.1 A Toller Diagram
\end{center}
}
$$
M_N(P_1,.,P_N)~\equiv ~
M_N\left(t_1,.,t_{N-3},g_1,.,g_{N-3}\right)
\auto\label{mng}
$$
where each $g_i$ is an element of the little group of $Q_i$. If $t_i >0$
the little group is $ SO(3)$ while for $t_i < 0$ it is 
$ SO(2,1)$. Taking $t_i > 0, ~\forall~ i~$, and using the parametrization 
$$
g_i=u_z(\mu_i)u_x(\theta_i)u_z(\nu_i) ~, ~~~~~0\leq \theta_i \leq \pi~,
~~0\leq \mu_i, \nu_i \leq 2\pi
\auto\label{par}
$$  
$M_N$ becomes a function of (N-3) $t_i$ variables, (N-3) $z_j$ variables,
where $z_i = \cos\theta_i ~,$ and (N-4) $u_{jk}$ variables, where
$u_{ij} = e^{i(\nu_j -\mu_i)}$. 

Multi-regge Limits are defined as 
$$
z_1, \ldots, z_{N-3} \longrightarrow \infty ~,
~\forall~ t_i, u_{ij} ~~fixed.
\auto
$$ 
These limits are physical when $t_i <0,~\forall~ i~$. Helicity-pole limits
in which some $z_i~\to \infty$ and some 
$u_{jk} \to \infty~$ (or $~0~$) involve fewer large invariants.
For any invariant $S_{mn\cdots
r}=\left(p_m+p_n+\ldots+p_r\right)^2$ the multi-regge limit gives the
asymptotic result
$$
S_{mn\ldots r }~~\centerunder{$\large\sim$}{
\raisebox{-2mm}{$z_j\to
\infty\ \forall_j$}}~~        
f(\til{t},\til{\omega})~z_{j_1}z_{j_2}
\cdots z_{j_s} \auto\label{3.48}
$$
where $j_1,\,j_2,\ldots,j_s$ denotes the {\it longest} path through
the Toller diagram linking any two of the external momenta contained in
$S_{mn\cdots r}$.

The angular variables can similarly be introduced when 
the external $p_i$ and the internal $Q_j$ are all spacelike. 
The analysis of (\ref{4.12})--(\ref{4.18}) then extends naturally to an 
$N$-point amplitude, as a function of the $z_j$-variables. 
From (\ref{4.15})--(\ref{4.18}), 
applying a complex boost to a real spacelike momentum
vector takes this vector into a tube of the form
(\ref{4.10}). More generally, if the imaginary part of the momentum vector
already satisfies (\ref{4.10}), and in addition the real part is a
spacelike vector, then both properties are preserved by a complex boost.
As a result the application of successive ``complex~$z_j$'' transformations, 
with the $\cosh\omega_{jk}$ and $t_j$ kept real, takes all
the momenta involved into a tube of the form (\ref{4.11}). Since 
$$
\eqalign{~~~~~~\cosh\Big[\beta_j+&
\beta_{j+1} +\cdots +\beta_{j+r}\Big] \cr
&\centerunder{$\large\sim$}{\raisebox{-3mm}
{$\scriptstyle z_j,z_{j+1},z_{j+2},\ldots\to\infty$}}\;
\cosh\beta_j\cosh\beta_{j+1}\ldots\cosh\beta_{j+r} 
~= ~ z_jz_{j+1}\ldots z_{j+r} }
\auto\label{4.20}
$$
analyticity in the tubes (\ref{4.11}) transfers
asymptotically into analyticity in the ``$z_j$-cones'' bounded by the
cuts
$$
{\rm Im}\left(~\prod_\Delta z_j~\right)~=~0\quad \forall~ \Delta,
\auto\label{4.21}
$$
where now $\Delta$ is any subset of $j=1,\ldots,N-3$ associated with {\it
adjacent} lines in the Toller diagram. This is the 
anticipated generalization of the off-shell cut-plane analyticity 
illustrated in Fig.~2.1(a).

Consider next the on mass-shell analyticity properties given in the 
previous Section. From (\ref{3.48}) all normal threshold cuts
$$
{\rm Im}\,S_{mn\ldots r}=0\qquad \forall\, mn\ldots r,\auto\label{4.22a}
$$
lie asymptotically within the cuts (\ref{4.21}). Also it can be shown 
that the assignment of imaginary parts giving a bad 
boundary-value is incompatible with (\ref{3.48}). A 
generalization of the situation depicted in Fig.~4.2 takes place. 
\begin{center}
\leavevmode
\epsfxsize=4in
\epsffile{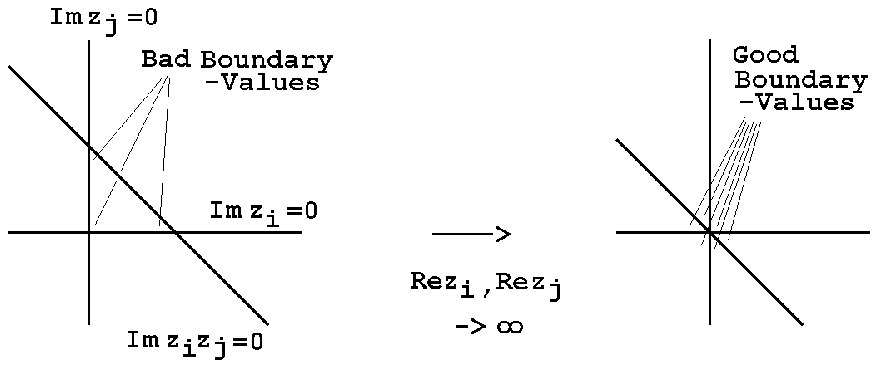}

Fig.~4.2 Asymptotic Cut Structure 
\end{center}
The bad boundary-values disappear asymptotically and only good functions
appear on either side of the asymptotic cuts (\ref{4.21}). Consequently
analyticity in the neighborhood of the cuts (\ref{4.21}) parallels that of
Fig.~2.1(b) and we conclude that a generalization of Fig.~2.1(c) also holds.

In the next Section we will disperse in 
the N-3 $z_j $ variables.
To apply the Bargman-Weil Theorem, the cuts (\ref{4.21}) 
must only intersect N-3 at a time. If the bad functions 
are chosen to satisfy the Steinmann relations, greater then
(N-3)-fold multiple discontinuities are zero and the theorem can be applied.
That the bad boundary-values become inaccessible asymptotically is, 
therefore, an essential pre-requisite.

\subsection{Multiparticle Asymptotic Dispersion Relations. }

Consider a function $f(\til{z})=f(z_1,\ldots,z_n)$ analytic
in ${\rm\bf C}^n$  minus a set of cuts
$c_m$ 
$$
c_m=\left\{\til{z}\in {\rm\bf C}^n;\quad {\rm Im}\,
s_m(\til{z})=0\right\}.\auto\label{4.28}
$$
If $I^\lambda$ is the intersection
of $n$ such cuts and $\Delta^\lambda(\til{z})$ is the multiple discontinuity 
$$
\Delta^\lambda(\til{z})=
\sum(-1)^{n'}~f\left(\til{z}\left({\rm Re}\,s_{\lambda_1}\pm
i\,0,\ldots,{\rm Re}\,s_{\lambda_n}\pm i\,0\right)\right),\qquad\lambda\equiv
(\lambda+1,\ldots,\lambda_n)
\auto\label{4.29}
$$
(the sum is over all combinations of $\pm$ signs and $n'$ is the number
of minus signs), the Bargman Weil Theorem says that we can write
$$
f(\til{z})~=~\sum_\lambda ~f^\lambda(\til{z})
+f^0(\til{z}),
\auto\label{4.30}
$$
where the sum is over all sets of $n$ cuts $\lambda$. $f^0$ includes
possible contributions from intersections of less than $n$ cuts together
with the ``sphere'' at infinity and
$$
f^\lambda(\til{z})={1\over (2\pi
i)^n}\int_{\til{z}'\in I^\lambda}dz'_1\ldots dz'_n~
\Delta^\lambda(\til{z}')\times\det
\left(\tild{q}^{\lambda_1},\
\tild{q}{}^{\lambda_2},\ldots \tild{q}
{}^{\lambda_n}\right).
\auto\label{4.31}
$$
The generalized dispersion denominators
$\tild{q}{}^{\lambda_m}$
must satisfy
$$
\tild{q}^{\lambda_m}(\til{z},\til{z}')\cdot (\til{z}'
-\til{z})=1.\auto\label{4.32}
$$
If we change variables
to the $s_{\lambda_m}$ and write (\ref{4.31}) in the form
$$                                
f^\lambda\left(\til{z}
(s_\lambda)\right)=     
{1\over (2\pi i)^n}
\int_{\til{\scriptstyle z}'\in I^\lambda} 
ds'_{\lambda_1}\ldots ds'_{\lambda_n}
\left({\partial \til{z}\over \partial s_\lambda}
\right)_{\til{\scriptstyle z}=\til{\scriptstyle z}'} 
\Delta^\lambda(s'_\lambda) ~
{D^\lambda(\til{z},\til{z}'
(s'_\lambda))\over 
(s'_{\lambda_1}-s_{\lambda_1}(\til{z}))\ldots
(s'_{\lambda_n}-s_{\lambda_n}(\til{z}))
} 
\auto\label{4.33}
$$
(\ref{4.32}) is satisfied if $D^\lambda(\til{z},\til{z}')$ is the
determinant of functions $p^\lambda_{m\ell}$ satisfying 
$$
s_{\lambda_m}(\til{z})-s_{\lambda_m}
(\til{z}')~=~\sum_\ell~ p^\lambda_{\ell
m}(\til{z},\til{z}')~
(z_\ell-z'_\ell).\auto\label{4.34}
$$
If the $s_{\lambda_m}$
are simple polynomials of the $z_j$ we can write                 
$$
D^\lambda(\til{z},\til{z}')=D^\lambda(\til{z},\til{z})+
\left[s_{\lambda_1}(\til{z})-s_{\lambda_1}(\til{z}')\right]
E^\lambda_1(\til{z},\til{z}')+\ldots +
\left[ s_{\lambda_n}(\til{z})-s_{\lambda_n}(\til{z}')\right]
E^\lambda_n(\til{z},\til{z}'),\auto\label{4.36}
$$
where the $E^\lambda_m(\til{z},\til{z}')$ are also polynomials.
Substituting (\ref{4.36}) into (\ref{4.33}) the first term gives the
simple expression
$$
f^\lambda\left(\til{z}(s_\lambda)\right)=
{1\over (2\pi i)^n}\int_{I^\lambda}ds'_{\lambda_1}\ldots ds'_{\lambda_n}
{\Delta^\lambda(s'_\lambda)\over
\left(s'_{\lambda_1}-s_{\lambda_m}(\til{z})\right)\ldots
\left(s'_{\lambda_n}-s_{\lambda_n}(\til{z})\right)},
\auto\label{4.37}
$$
while the remaining terms in (\ref{4.36}) cancel at least one denominator
in (\ref{4.33}) and so can be included in the $f^0$ term appearing in
(\ref{4.30}).

The multi-Regge behavior which generalizes (\ref{4.4})
is that
$$                         
f(\til{z})\equiv                  
M(\til{t},\til{z},\til{u})\,~~\centerunder{$\large\sim$}{
\raisebox{-2mm}{$\scriptstyle z_1,\ldots,z_{N-3}\to \infty$}}~~\,
\prod_{j=1}^{N-3}~(z_j)^{\alpha_j(t_j)}
\auto\label{4.38}
$$
(where again there may also be $\ln z_j $ factors). Since this
represents a function with $n\equiv (N-3)$ cuts 
it can only originate from the
$f^\lambda(\til{z})$ terms in (\ref{4.30}). Also 
non-leading functions
of the $z_j$ appearing in the $s_{\lambda_m}(\til{z})$
can be dropped and their effects absorbed in $f^0$. Consequently we can 
write an asymptotic dispersion relation
$$
M_N~=~\sum_{C}~M_N^C ~+~M^0
\auto\label{adr}
$$
where the $\sum_{C}$ is over all sets of (N-3) Regge limit asymptotic cuts
and the asymptotic form (\ref{3.48}) expressing all
invariants as polynomials in the $z_j$ justifies writing each of
the $M_N^C$ in the form
$$
M_N^C ~=~ {1\over (2\pi i)^{N-3}}
\int \frac{ ds'_1\ldots ds'_{N-3}~\Delta^C(
..t_i.,..u_{jk}.,..s'_i.)}
{(s'_1-s_1)(s'_2-s_2)\ldots (s'_{N-3}-s_{N-3})} 
\auto\label{mnc}
$$
Therefore to write a full asymptotic dispersion relation we need only
to enumerate the complete set of multiple discontinuities 
(\ref{4.29}). A complete discussion of the asymptotic dispersion relation
corresponding to the Toller diagram of Fig.~4.1 has been given (Stapp and 
White, 1982).

\subsection{Classification of Multiple Discontinuities}

Given that the Steinmann relations are satisfied, 
multiple discontinuities can be counted 
and classified using ``hexagraphs''   (White, 1976,1991).  The 
hexagraphs\footnote{There must
be a close relationship between hexagraph amplitudes and the
field theory GRFs, but so far this has not been studied.} 
associated with a Toller diagram are obtained by
redrawing the diagram in all possible ways (in a plane) with the
internal lines drawn as horizontal lines and the internal vertices drawn
separately, with relative angles of $120^o$, and joined to the horizontal
lines. A hexagraph associated with the Toller diagram of Fig.~4.1 is shown
in Fig.~4.3 
(a). (The $j$ and $n$ labels are explained 
below.) 
\begin{center}
\leavevmode
\epsfxsize=4.5in
\epsffile{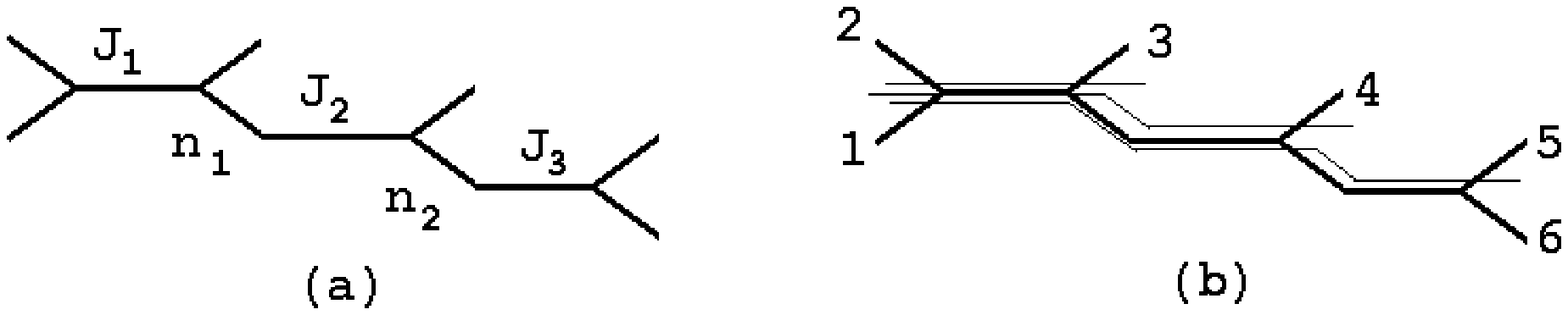}

Fig.~4.3 (a) A Hexagraph from Fig.~4.1, (b) Allowable Cuts.
\end{center}
The multiple discontinuities associated with a particular hexagraph all
appear in the same $s$-channel physical region - obtained 
by regarding the 
scattering particles as entering from the bottom and
exiting at the top. Each graph is also associated with a particular $t$-channel 
- obtained by regarding the 
scattering particles as entering from the left and
exiting to the right. 

An ``allowable discontinuity'' of a hexagraph is a cut drawn through
internal lines that connects the particles involved in the discontinuity
channel, and enters and exits only between non-horizontal lines. A set of
allowable discontinuities of the hexagraph of Fig.~4.3(a) is shown in
Fig.~4.3(b). For an N-point amplitude, a hexagraph represents all sets of
(N-3) asymptotically independent cuts that are allowable discontinuities and
satisfy the Steinmann relations. Each multiple discontinuity appears in only
one hexagraph.

\section{Multi-Regge Theory.}

\setcounter{equation}{0}

The break-up of an $N$-point amplitude into hexagraph amplitudes
allows multi-regge theory to be 
developed as a generalization of elementary 
regge theory. In this Section we briefly describe the central elements. (For 
a full development see White, 1976,1991,1998,1999).

\subsection{Partial-Wave Expansions}

For a function $f(g)$ defined on SO(3) we can write $$ 
f(g)=\sum^\infty_{J=0}\,\sum_{|n|,|n'|<J}D^J_{nn'}(g)a_{J nn'},
\auto\label{pw}
$$
where the $D^J_{nn'}(g)$ are representation functions. 
In a $t$-channel where all the little groups are SO(3), 
(\ref{mng}) gives the generalized partial-wave expansion 
$$
\eqalign{
M_N(\til{t},g_1,\ldots,g_{N-3})={}&\sum^\infty_{J_1=0}\,
\sum_{|n_1|,|n'_1|<J_1}\ldots\sum^\infty_{J_{N-3}=0}~~\,
\sum^\infty_{|n_{N-3}|,|n'_{N-3}|<J_{N-3}}\cr
&D^{J_1}_{n_1n'_1}(g_1)\ldots D^{J_{N-3}}_{n_{N-3},n'_{N-3}}(g_{N-3})
a_{J_1,n_1,n'_1,\ldots,J_{N-3},n_{N-3},n'_{N-3}}(\til{t}).}
\auto\label{pw1}
$$
Because $M_N$ depends only on particular combinations of the 
azimuthal angles, the helicity labels in (\ref{pw1}) are also 
constrained. For a particular hexagraph the $j$ variables can be associated
with the horizontal lines, while independent $n$ and $n'$ variables are
associated with sloping lines, as in Fig.~4.3(a). 

\subsection{Froissart-Gribov Continuations.}

Since a hexagraph amplitude $M^H$ has simultaneous cuts in only 
N-3 large invariants, a Sommerfeld-Watson representation 
that reproduces this cut structure
is obtained by transforming only N-3 of the 
angular-momentum and helicity sums in (\ref{pw1}). Correspondingly, 
Froisart-Gribov continuations can be made only in the complex planes
of the relevant indices. The hexagraph groups together
those sets of cuts for which continuations in the same helicity and
angular momentum variables can be made. The essential feature
is that the helicity
labels, which are attached to sloping lines of the hexagraph, are always
coupled to (that is differ only by an integer from) the angular momentum
associated with the corresponding horizontal line. For example, in 
Fig.~4.3(a) $n_1$ would be coupled to $j_2$ and $n_2$ would be coupled to 
$j_3$. Signature is introduced 
by combining together all hexagraphs that are related by twists about
horizontal lines and so have the same $t$-channel.

\subsection{Sommerfeld-Watson Representations }

The form of the Sommerfeld-Watson representation for 
a set of hexagraphs with the same $t$-channel is directly
related to the rules for Froisart-Gribov continuations. 
For example, the representation of the hexagraph of Fig.~4.3(a) has the form
$$                
\eqalign{                                  
A_H={}{1\over 8}\int_{C_{n_2}}
{dn_2\,u_2^{n_2}\over \sin\pi
n_2}~&\int_{C_{n_1}} {dn_1\,u_1^{n_1}\over
\sin\pi(n_1-n_2)}~\int_{C_{J_1}} {dJ_1~d^{J_1}_{0,n_1}(z_1)\over
\sin\pi(J_1-n_1)}\cr
&\times~   \sum^\infty_{{\scriptstyle J_2-n_1=N_1=0\atop
\scriptstyle J_3-n_2=N_2=0}}
d^{J_2}_{n_1,n_2}(z_2)d^{J_3}_{n_2,0}(z_3)a_{N_2N_3}(J_1,n_1,n_2,
\til{t}) ~~ + ~~\cdots
,}\auto\label{5.45}
$$
Such representations give the asymptotic behaviour in both
multi-Regge and helicity-pole limits. In the multi-regge limit
the contribution of regge poles at 
$j_1 = \alpha_1(t_1)$, $j_1 = \alpha_1(t_1)$, and $j_1 = \alpha_1(t_1)$ 
in (\ref{5.45}) gives
$$  
\eqalign{A_H  ~  \centerunder{\centerunder{ 
\centerunder{$ \sim$}{\raisebox{-2mm}{$\scriptstyle
z_1\to\infty $}}}{\raisebox{-2mm}{$\scriptstyle z_2\to\infty$}}}{
\raisebox{-2mm}{$\scriptstyle z_3\to\infty $}}~
&{z_1^{\alpha_1}z_2^{\alpha_2}z_3^{\alpha_3}\over 
\sin\pi\alpha_3\sin\pi(\alpha_2-\alpha_3)\sin\pi(\alpha_1-\alpha_2)}
\sum^\infty_{N_1=N_2=0}\biggl[\beta_{N_1N_2}^{\alpha_1\alpha_2\alpha_3}
u_1^{\alpha_2-N_1}u_2^{\alpha_3-N_2} \cr
&+ \beta_{N_1N_2}^{\alpha_1 -\alpha_2\alpha_3}
u_1^{- \alpha_2-N_1}u_2^{\alpha_3-N_2} + 
\beta_{N_1N_2}^{\alpha_1\alpha_2-\alpha_3}
u_1^{\alpha_2-N_1}u_2^{-\alpha_3-N_2} + 
\beta_{N_1N_2}^{\alpha_1-\alpha_2-\alpha_3}
u_1^{-\alpha_2-N_1}u_2^{-\alpha_3-N_2}\biggr]. }
\auto\label{mrll}
$$
In a helicity-pole limit, only the leading term appears.

\subsection{$t$-channel Unitarity in the $J$-plane}

The discontinuity across the $n$-particle threshold in any $t$-channel is 
shown in Fig.~5.1. 
\begin{center} 
\leavevmode
\epsfxsize=4in
\epsffile{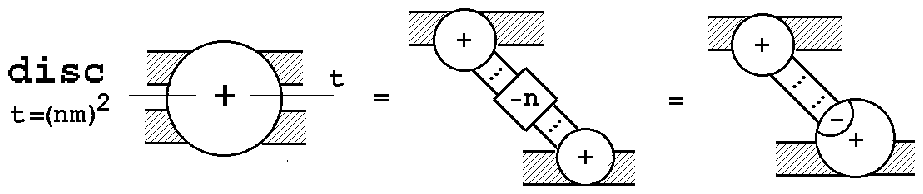}

Fig.~5.1 The $n$-particle Discontinuity.
\end{center}
The $n$-particle phase-space can be written in terms of group
variables as
$$
I_n(t)=i\int d\rho(t,t_1,\ldots )\int dg_L\prod_j
dg_j
\auto\label{nps}
$$
This integration is diagonalized by the partial-wave projection
(\ref{pw1}). It can also be shown that for regge behavior there
is a form of ``hexagraph diagonalization'' (White, 1991). 
For a hexagraph amplitude $A^H$ the
corresponding part of the unitarity integral can then be continued in the
complex $j$-plane in the form 
$$ 
a^{H+}_{\cdots J\cdots} - a^{H-}_{\cdots J\cdots} = 
i\int d\rho \sum_{\til{N}} 
\int {dn_1 dn_2  \over 
\sin\pi(J-n_1-n_2) }
\int {dn_3 dn_4 \over \sin\pi(n_1 -n_3 -n_4)} ~\cdots 
~a^{H_L +}_{\cdots J\til{N}
\til{n}}a^{H_R -}_{J\til{N}\til{n}\cdots} 
\auto\label{tcu}
$$
where $H_L$ and $H_R$ are hexagraphs whose ``product'' gives the hexagraph 
$H$ (White, 1991,1998). 

\subsection{Reggeon Unitarity }

If the amplitude $a^{H_L}$ in (\ref{tcu}) contains regge poles at
$n_i=\alpha_i$ then multi-reggeon thresholds (regge cuts) are generated by
the phase-space $\int d\rho $ together with the ``nonsense poles'' at $J=
n_1 +n_2 -1, n_1=n_3 + n_4 -1 , ~...~$. Corresponding angular momentum plane 
discontinuity formula can be derived from (\ref{tcu})
(in conjuction with sub-channel discontinuity formulae).
In general, from analagous unitarity equations, it can be shown 
that in any $J$-plane of 
any partial-wave amplitude, the threshold 
discontinuity due to $M$ Regge poles with trajectories 
\newline $\til{\alpha}= (\alpha_1, \alpha_2, \cdots \alpha_M)$
is given by the reggeon unitarity equation 
$$ 
\centerunder{disc}{\raisebox{-3mm}{$\scriptstyle J=\alpha_M(t)$}}~~ 
a^{H}_{\cdots J\cdots} 
~=~ {\xi}_{M} \int d\hat\rho~
a^{H_L+}_{\cdots J \til{\alpha}}
~a^{H_R-}_{J \til{\alpha} \cdots} 
~~{\delta\left(J-1-\sum^M_{k=1}
(\alpha_k-1)\right)\over \sin{\pi\over 2}(\alpha_1-\tau'_1)\ldots\sin{\pi\over
2}(\alpha_M-\tau'_M) }
\auto\label{rgu}
$$
where ${\xi}_{M}$ is a complicated signature factor and 
$a^{H_L+}_{\cdots J \til{\alpha}}$ and $a^{H_R- }_{J \til{\alpha}\cdots}$
are regge pole residues.
If $\alpha_i=\alpha(t_i)~\forall ~i~$, then 
the trajectory of the $M$-reggeon cut is 
$$
J~=~\alpha_M(t)~= ~M~\alpha(t/M^2) - M +1
\auto\label{mtr}
$$
Writing $t_i=k_i^2~~$ and using 
$$
\int dt_1 dt_2 \lambda^{-1/2}(t,t_1,t_2) ~= ~2\int d^2 k_1 d^2 k_2 \delta^2(k 
- k_1 - k_2) 
\auto\label{kperp}
$$ 
the phase-space integration
$\int d \hat{\rho} $ in (\ref{rgu}) can be written in terms of
two dimensional ``$~k_{\perp}$'' integrations, 
anticipating the reggeon diagram results of 
the direct $s$-channel QCD calculations~
that will be discussed in Section 7. 

The reggeon unitarity equations were
first derived in the mid-sixties (Gribov et al., 1965). However, there were
many uncertainties in the derivation before the development of asymptotic
dispersion relations as a fundamental basis for multi-regge theory. The
generality of reggeon unitarity makes it 
extremely powerful, particularly when applied to the problem of constructing 
the regge region QCD S-Matrix, as we discuss briefly in Section 7. First we 
discuss the abstract solution of these equations using Reggeon Field Theory 
(RFT).

\section{Reggeon Field Theory.}

\setcounter{equation}{0}

RFT has been derived and formulated from many different
starting points since Gribov's seminal work (Gribov, 1967,1968; for an early 
review of RFT see Abarbanel,
Bronzan et al., 1975). From an 
S-Matrix viewpoint RFT simply provides a solution of the reggeon
unitarity equations. Consider an isolated even-signature
pomeron regge pole with trajectory $j = \alpha(t)$ and the associated 
multi-pomeron cuts. When $\alpha(0) = 1$ 
(corresponding to the ``maximum strength'' postulate 
[7] of Section 2) then also $\alpha_M(0) = 1$. Therefore, 
all the multipomeron singularities accumulate at one point and 
a simultaneous solution of all the discontinuity formulae must be found.
Remarkably, a renormalization group formalism can be applied and a 
fixed-point solution found (Migdal et al., 1974; Abarbanel and Bronzan, 1974).

\subsection{Pomeron Phase-Space and the Effective Lagrangian.}

For the pomeron the signature factor $\xi_M \sim (-1)^{M-1}$, while 
the other signature factors in (\ref{rgu}) can be neglected. 
The $\delta$-function in (\ref{rgu}) becomes ``energy
conservation'' for the RFT energy $E=1-J$ and
a general solution is obtained by writing a
(non-relativistic) graphical expansion involving pomeron propagators and
vertices. If $\bar\phi(\til{x},y)$ and 
$\phi(\til{x},y)$ are respectively Pomeron creation and destruction 
operators ($\til{x}$ and $y$ are  conjugate to $k_{\perp}$ and $E$)
the corresponding effective lagrangian is
$$
\eqalign{
{\cal L}(\bar\phi,\phi){}=  {1\over 2}i\bar\phi
\centerover{${\partial\over \partial y}$}{$\leftrightarrow$}\phi-
&\alpha'_0\nabla \bar\phi\cdot\nabla\phi - \Delta_0\bar\phi\phi + 
\alpha''_0\nabla^2\bar\phi\cdot\nabla^2\phi \cdots 
- {i\over 2}\bigl[r_0\bar\phi^2\phi + r_0\bar\phi\phi^2 \cr
& + r_{01} \bar\phi\phi\nabla^2\phi+\cdots\bigr]
+{1\over 6}\left[\lambda_0\bar\phi^3\phi + \lambda_0\bar\phi\phi^3+\cdots
\right]+\cdots.}
\auto\label{7.13}
$$

\subsection{The Critical Pomeron}

If $|E|$, $|\kbar^2| < \mu $ initially, integrating out regions 
of $E$ and $\kbar$ so that $\mu\to \mu/\zeta$ gives a
renormalization group transformation which rescales each parameter $g$ in
the lagrangian (\ref{7.13}) by $g\to \zeta^{\nu_g} g+ g'$, where $\nu_g$ is
the canonical dimension of $g$. Bare parameters with $\nu_g<0$ are
``irrelevant'' and can be dropped. Since 
$$
\nu_{\Delta_0}=1,\quad \nu_{r_0}={1\over 2}\quad \nu_{\alpha'_0}=0
\auto\label{7.17}
$$
these parameters are kept. Expanding in powers of $\epsilon=4-D$, 
where $D$ is the number
of transverse dimensions, a fixed-point 
is found at 
$$
r^2={4\pi^2\over 3} \epsilon,\quad \Delta= 1- \alpha(0)= 0
\auto\label{7.18}
$$
Provided this fixed-point persists to $\epsilon=2$, an interacting pomeron 
theory satisfying (\ref{rgu}) exists that 
has the ``universality'' property familiar from critical 
phenomena. Consequently the asymptotic behavior 
can be calculated without knowledge of the initial bare
parameters. 

At leading-order in $\epsilon$ the elastic differential cross-section has
the scaling behavior 
$$
{d\sigma\over dt}\sim (\ln s)^{\epsilon/6}F^2\left(
-{\alpha'_0 t[\ln s]^{1+\epsilon/24}\over K}\right)\auto\label{7.45}
$$
where (Abarbanel, Bartels et al., 1975) 
$$
F(x)=~x^{(-\epsilon/12)/(1+\epsilon/24)}\Gamma\left(1+\epsilon/12\right)~
\int^{+i\infty}_{-i\infty}
{dw\,e^{-wx^{1/(1+\epsilon/24)}}\left[1+\bar\eta/2\right]^{-\epsilon/12}\over
(-w)^{1+\epsilon/12}\left[1+\bar\eta(1+\epsilon/24)\right]\left[1+ \bar\eta 
/2 \right]}~,
\auto\label{7.46}
$$
with
$$
\bar\eta(1+\bar\eta/2)^{\epsilon /24}=(-w)^{-1-\epsilon/24}~,~~~
K=\left[{(8\pi)^2\epsilon(\alpha'_0)^{4-\epsilon}/2\over 6r^2_0}\right]
^{1/12}. 
\auto\label{7.460}
$$
Setting $t=0$ in (\ref{7.45}) leads to the well-known
Critical Pomeron result for the total cross-section. 
A scaling law  similar to (\ref{7.45}), but even more elaborate, can also 
be derived for the triple-regge region of
the one-particle inclusive cross-section (Abarbanel, Bartels et al., 1975,
other Critical Pomeron scaling
properties are reviewed in Moshe, 1978).

In a sense the Critical Pomeron is the summit of abstract S-Matrix Theory.
It satisfies all known unitarity constraints on a theory of rising
cross-sections and it is unique in this respect. 
It has been formulated without reference to any underlying theory and 
provides a uniquely
attractive possibility for the high-energy behavior of an S-Matrix 
satisfying the maximum strength postulate. 

\subsection{The Super-Critical Pomeron}

The nature of the ``supercritical phase'' that appears when the pomeron 
intercept is pushed beyond the critical point was much debated in the 
mid-seventies. An ``expanding disc'' solution that, unfortunately, does 
not satisfy reggeon unitarity was proposed by several authors 
(see, for example, Amati et al., 1975).
The efffective lagrangian close to the critical point is given by 
(\ref{7.13}) with only $\alpha_0',~ \Delta_0 $ and $r_0~ \neq 0$.
A supercritical theory that has $\Delta_0 < 0$ and does satisfy reggeon 
unitarity can be defined
by using the stationary point at 
$$
\phi ~=~ \bar\phi~ = ~2i\Delta_0/3r_0
\auto\label{ev}
$$ 
to introduce a ``pomeron
condensate'' (White, 1991). The condensate generates
new classes of RFT diagrams whose physical interpretation is subtle. 
Reggeon unitarity determines that the $k_{\perp}$ poles produced by
zero energy two-pomeron propagators are to be interpreted as due to 
particle poles lying on an odd-signature 
trajectory degenerate with that of the pomeron. The odd-signature reggeon 
couples pairwise to the pomeron. 
A general characterization of the supercritical phase introduced this way 
is, therefore, that the divergences in rapidity produced by $\Delta_0 < 0$
are converted to vector particle divergences in $k_{\perp}$. The divergences
are then associated with the ``deconfinement of a vector particle'' on the
pomeron trajectory. 

An obviously important question is whether the super-critical 
phase can be realized in QCD? The appearance of a 
a reggeized vector particle (a ``gluon'') strongly suggests the spontaneous
breaking of a gauge symmetry that would correspond to a color superconducting 
phase of QCD.

\section{QCD and the Critical Pomeron}

\setcounter{equation}{0}

\subsection{Reggeon Diagrams in QCD. }

Leading-log Regge limit calculations of elastic and multi-regge production 
amplitudes in (spontaneously-broken) gauge theories
show that both gluons and quarks lie on Regge trajectories
(Fadin et al., 1977; Cheng and Lo, 1976,1977; 
Fadin and Sherman, 1978). Non-leading log calculations are described
by reggeon diagrams involving reggeized gluons and
quarks, just as required by reggeon unitarity.
Gluon reggeon diagrams involve a reggeon propagator for each multi-reggeon 
state and also gluon particle poles e.g. 
$$
\hbox{two-reggeon state} ~~~\leftrightarrow ~~
\int {d^2k_1 \over (k_1^2 +M^2) } {d^2k_2 \over (k_2^2 + M^2)}~~ 
{\delta^2(k_1'+k_2'-k_1-k_2)
\over J-1 +  \Delta(k_1^2,M^2) + \Delta(k_2^2,M^2)}
\auto\label{2rs}
$$
To leading order in the gauge coupling $g^2$ this gives (\ref{rgu}) with the 
particle poles producing the additional signature factors associated with an 
odd-signature reggeon. Bronzan and Sugar showed
that the two-two reggeon interaction 
$$                        
\Gamma_{22}(\kbar_1,\kbar_2,\kbar_1',\kbar_2',M^2)~= ~g^2~~
{{(\kbar^2_1+M^2)(
{\kbar^2_2}'+M^2)+(\kbar^2_2+M^2)(
{\kbar^2_1}'+M^2)}\over
{(\kbar_1-\kbar_1')^2+M^2}} ~~+ ~~\cdots
\auto\label{rgi}
$$
could be extracted from sixth-order calculations and used to predict the 
independently calculated eighth and tenth orders 
(Bronzan and Sugar, 1978). The form of the interaction 
(\ref{rgi}) can also be deduced
directly from $t$-chhannel unitarity (White, 1994).
The well-known BFKL equation 
is obtained by summing to all orders the interaction 
(\ref{rgi}) for the state 
(\ref{2rs}) and taking the limit $M^2 \to 0$
(Fadin et al., 1977; Balitsky and Lipatov, 1978).

When the symmetry breaking (producing the mass $M$) is due to 
color triplet scalars the 
theory can be formulated gauge-invariantly (Banks and Rabinovici, 1979; 't 
Hooft, 1980) and, as discussed earlier, all of the
analyticity properties of field theory and S-Matrix Theory that have been 
used in deriving reggeon unitarity apply perturbatively. Therefore, to all 
orders, the regge limit should be described by gluon, and quark, reggeon
diagrams. 

When the gluon mass $M \to 0$ there are infra-red divergences
(due to the particle poles) in the reggeon states, in the interactions,  and
in the trajectory function. These  divergences combine to 
exponentiate to zero all
reggeon diagrams that do not carry zero color in the $t$-channel. In 
color-zero channels the divergences cancel. This
is not confinement, however, because color-zero multi-gluon singularities
are still present. 

\subsection{Color Superconductivity and the Supercritical Pomeron}

Gluon reggeon diagrams differ from RFT pomeron diagrams only because of
the gluon particle poles. The presence of these poles has several
consequences. In particular, it allows all 
gluon reggeon interactions to have the same scaling dimension under the 
renormalization group transformation discussed in the previous Section.
Hence finding a fixed point for reggeized gluon interactions would be very 
difficult, if not impossible. To apply the renormalization group 
a (confinement) mechanism has to be found whereby the particle poles disappear.
A partial mechanism could be provided by the reverse of the RFT
phase-transition to the supercritical pomeron. As described above, in
this transition, ``gluon 
poles'' appear from within pomeron diagrams. Before this
possibility can be considered, however, it is first necessary to 
identify the supercritical RFT phase within gauge theory reggeon diagrams. 

The defining features of the supercritical phase, i.e. a regge pole pomeron
with a reggeized massive vector partner and a ``pomeron condensate'', 
could be realized in a color superconducting phase of QCD in which SU(3)
color is broken to SU(2). The symmetry-breaking vector mass would then be
identified with the RFT order parameter and the Critical Pomeron would
appear at the critical point where color superconductivity disappears and
full SU(3) symmetry is restored. For this identification to be made there
should also be a reggeon condensate with the quantum numbers of the
winding-number current, associated (presumably) with spectral flow of
the Dirac sea. Recent results show that quark loops do indeed produce
anomalous reggeon interactions that could produce this condensate via
infra-red divergences (White, 1999). The hope is that, within the reggeon
diagrams that describe the fully superconducting phase of QCD (in which the
gauge symmetry is completely broken), partial restoration of the gauge
symmetry produces infra-red divergences involving the anomaly that lead to
the appearance of the supercritical RFT phase. Restoration of the full SU(3)
gauge symmetry would then give the Critical Pomeron. 

\subsection{Quark Saturation and an Infra-Red Fixed Point}

In general, a gauge-invariant cut-off must be present if
a gauge symmetry is to be restored smoothly. In the regge limit a 
$k_{\perp}$ cut-off can be used. However, it then becomes an additional
relevant parameter for the RFT phase transition. For Critical Pomeron
behavior to be present after the cut-off is removed 
and the short-distance part of the 
theory is included, it must be that a smooth
parameter variation can introduce the massive vector of the supercritical
phase. This can be done via the Higgs mechanism only if 
the Higgs self-coupling is also asymptotically-free. This is the case only 
when QCD is ``flavor-saturated'', i.e. the maximum number of flavors allowed by 
asymptotic freedom is present. When all quarks are 
massless flavor saturation also produces 
an infra-red fixed-point for the gauge coupling
- a property that is closely related to the 
presence of an RFT fixed-point (White, 1993).

Flavor saturation is produced by sixteen conventional quark flavors 
but this is physically unrealistic. A second possibility 
is six conventional quark flavors plus two flavors of color sextet quarks.
This may very well be physically realistic. 
If a doublet of color sextet quarks exists, with conventional electroweak 
quantum numbers, the sextet pions produced by the breaking of the 
sextet chiral 
symmetry can provide the electroweak ``Higgs sector''. That is 
QCD chiral symmetry breaking in the sextet sector will 
simultaneously break 
the electroweak gauge symmetry. This gives the attractive possibility 
that the electroweak scale is actually a QCD chiral scale, rather than a new
scale produced by new physics beyond the Standard Model.

The sextet doublet may even 
be necessary for the self-consistency of QCD. In addition to the sextet 
pions that produce the longitudinal components of the massive $W^{\pm}$ and 
$Z^0$, the sextet doublet also produces an axion that can
prevent $CP$ violation within QCD. Apart from it's being unobserved 
experimentally, ``Strong $CP$ violation'' is undesirabale from the present 
perspective because it 
would destroy the crucial even signature property of the pomeron that allows 
a fixed-point solution of reggeon unitarity.

\subsection{Uniqueness of the S-Matrix ?} 

From the discussion of the last subsection it appears 
that QCD with a fixed quark content may 
uniquely produce the Critical Pomeron. It may also be possible to argue 
that the graphs of the supercritical phase, when 
studied in detail, uniquely correspond to color superconducting QCD with the 
gauge group fixed to be SU(3). If the Critical Pomeron is the only
high-energy solution of unitarity that can match with asymptotic
freedom then perhaps there is a uniqueness property for the
strong-interaction S-Matrix close to that conjectured by the early 
S-Matrix enthusiasts. The only difference would be that the mass scales
involved are not determined, only the underlying massless theory. 

But, why should Critical Pomeron asymptotic behavior be unique? 
Why, in particular, should the pomeron be only a single regge 
pole plus multipomeron cuts? That regge poles and the regge cuts built out 
of them are the only angular momentum plane singularities 
is almost certainly required for 
the solution of $t$-channel unitarity. A 
single regge pole pomeron uniquely has the factorization 
properties needed to be associated with a universal wee-parton  distribution 
in hadrons. This universality property allows properties of a
non-trivial vacuum to be carried by wee-partons. This, in turn, allows
the most powerful form of the parton model and the maximal 
applicability of short-distance perturbation theory. Such properties may 
well be essential to produce a completely finite (and unitary) S-Matrix.

So it is indeed conceivable that the underlying massless 
field theory giving all the 
desired properties of the strong-interaction S-Matrix interaction is unique. 
The mass scales involved are presumably determined by the unification with 
the electroweak interaction. Could the full S-Matrix including 
the electroweak interaction be unique? To produce full asymptotic 
freedom at short-distances there should be an underlying non-abelian gauge 
group (this will also reggeize the photon - a necessary ingredient 
for the definition of an S-Matrix). The gauge symmetry should be 
partially-broken since any gauge symmetry larger than SU(3) 
produces a more complicated 
structure of pomeron regge poles. To
produce the saturation of QCD needed to obtain the Critical Pomeron, the
underlying gauge group has to be close to saturation. If this gauge symmetry
is left-handed (which, of course, may not be necessary) 
then anomaly cancelation implies that 
the possibilities are indeed extremely limited (Kang and White, 1987).
It is, therefore, not unreasonable to conjecture that a full set of 
consistency constraints including, presumably, some not yet formulated, do
lead to a unique S-Matrix. 

If the uniqueness of the S-Matrix determines the underlying 
gauge theory, before the existence of gravity is considered, this would be 
strongly counter to today's prevailing philosophy. Even though uniqueness is 
indeed envisaged after the inclusion of gravity.

\newpage

\noindent {\Large {\bf References}} 

\vspace{0.2in}

{\parskip 3pt 
 
\noindent Abarbanel H D I and Bronzan J B (1974) 
{\it Phys. Rev.} {\bf D9}, 2397.

\noindent Abarbanel H D I, Bartels J, Bronzan  J B and Sidhu D (1975)
{\it Phys. Rev} {\bf D12}, 2798.

\noindent Abarbanel H D I, Bronzan J B, Sugar R L and White A R (1975)
{\it Phys. Repts.} {\bf 21C}, 119. 

\noindent Amati D, Ciafaloni M, LeBellac M and Marchesini G (1975) {\it 
Nucl. Phys.} {\bf B115}, 107.

\noindent Balitsky Ya Ya and Lipatov L N (1978) {\it Sov. J.
Nucl. Phys.} {\bf 28}, 822.

\noindent Banks T and Rabinovici E (1979) {\it Nucl. Phys.} {\bf B160}, 349. 

\noindent Bardeen W A, Fritzsch H and Gell-Mann M (1973) Light-Cone Current 
Algebra, $\pi_o$ Decay and $e^+e^-$ Annihilation. In Gatto R (ed)
{\it Scale and Conformal Symmetry in Hadron Physics}. Wiley \& Sons. 

\noindent Bartels J (1993) {\it Z. Phys.} {\bf C60}, 471. 

\noindent Bronzan J B and Sugar R L (1978) {\it Phys. Rev.} {\bf D17}, 585. 

\noindent Bros J and Lassalle M (1976) An Approach to the Non-Linear Program 
of General Quantum Field Theory. In Balian R and Iagolnitzer D (eds) 
{\it Structural Analysis of Collision Amplitudes}, pp 427-503. North Holland.

\noindent Cahill K E and Stapp H P (1975) {\it Ann. Phys.} {\bf 90}, 438.

\noindent Chandler C and Stapp H P (1969) {\it J. Math. Phys.} {\bf 10}, 826. 

\noindent Cheng H and Lo C Y (1976) Phys. Rev. {\bf D13}, 1131.

\noindent Cheng H and Lo C Y (1977) Phys. Rev. {\bf D15}, 2959 (1977). 

\noindent Chew G F (1962) {\it S-Matrix Theory of Strong Interactions}.
Benjamin. 

\noindent Chew G F and Mandelstam S (1960) {\it Phys. Rev.} {\bf 119}, 467.

\noindent Chew G F, Frautschi S C and Mandelstam S (1962) {\it Phys. Rev.} 
{\bf 126}, 1202. 

\noindent Coster J and Stapp H P (1975) {\it J. Math. Phys.} {\bf 16}, 1288.

\noindent De~Rujala A, Georgi H and Glashow S (1975) {\it Phys. Rev.}
{\bf D12}, 147. 

\noindent Coriano C and White A R (1996) {\it Nucl. Phys.} {\bf B468}, 175.

\noindent Dolen R, Horn D and Schmid C (1968) {\it Phys. Rev.} {\bf 1666},
1768.

\noindent Dorey P (1998) {\it Lectures on Exact S-Matrices.} hep-th/9810026. 

\noindent Eden R J, Landshoff P V, Olive D I and Polkinghorne J C (1966) 
{\it The Analytic S-Matrix}. Cambridge University. 

\noindent Epstein H (1965)  Axiomatic Field Theory. In  {\it
Proc. Brandeis Summer Institute.} Gordon and Breach. 

\noindent Epstein H, Glaser V and Stora R (1976) General Properties of the
n-Point Functions in Local Quantun Field Theory. In Balian R and
Iagolnitzer D (eds) {\it Structural Analysis of Collision Amplitudes}, pp
5-93. North Holland. 

\noindent Fadin V S and Lipatov L N (1996) {\it Nucl. Phys.} {\bf B477}, 767.

\noindent Fadin V S and Sherman V E (1978) {\it Sov. Phys. JETP} {\bf 45}, 861.

\noindent Fadin V S, Kuraev A E and Lipatov L N (1977) {\it Sov. Phys.
JETP} {\bf 45}, 199.

\noindent Feynman R P (1961) Comment. In Stoops R (ed) {\it The Quantum Theory 
of Fields, the 12th Solvay Conference}, p177. Interscience. 

\noindent Feynman R P (1967) Field Theory as a Guide to Strong 
Interactions. In Hagen C R, Guralnik G and Mathur V A (eds) 
{\it Proc. 1967 International
Conference on Particles and Fields}, pp 111-127.  Interscience. 

\noindent Fritzsch H and Gell-Mann M (1972) Current Algebra, Quarks and 
What Else? {\it Proc. XVI International 
Conference on High-Energy Physics}, Vol. 2 pp 135-165. NAL. 

\noindent Gell-Mann M (1962) {\it Phys. Rev.} {\bf 125}, 1067. 

\noindent Goldberger M L (1961) Theory and Applications of Single Variable 
Dispersion Relations.  In Stoops R (ed.) {\it The Quantum Theory 
of Fields, the 12th Solvay Conference}, p177. Interscience. 

\noindent Gribov V N (1962) Partial Waves with Complex Angular Momenta and 
their Moving Singularities. In {\it Proc. 1962 International 
Conference on High Energy Physics}. CERN.

\noindent Gribov V N (1967) Cuts in the Angular Momentum Plane and High 
Energy Asymptotic Behavior. In Hagen C R, Guralnik G and Mathur V A (eds) 
{\it Proc. 1967 International
Conference on Particles and Fields}, pp 621-631.  Interscience. 

\noindent Gribov V N (1968) {\it Soviet Phys. JETP } {\bf 26}, 414. 

\noindent Gribov V N (1969) Lectures on the  
Theory of Complex Angular Momenta. Leningrad Physical-Technical Institute 
- to be published.

\noindent Gribov V N, Pomeranchuk I Ya and Ter-Martirosyan K A (1965)
{\it Phys. Rev.} {\bf 139B}, 184.

\noindent Grisaru M T and Schnitzer H J (1979) Phys. Rev. {\bf D20}, 784.

\noindent Gross D J (1999) {\it Nucl. Phys. Proc. Suppl.} {\bf 74}, 426.

\noindent Gross D J and Wilczek F (1973) {\it Phys. Rev.} {\bf D8}, 3633.

\noindent Gross D J and Wilczek F (1974) {\it Phys. Rev.} {\bf D9}, 980. 

\noindent Gunson J (1965) {\it J. Math. Phys.} {\bf 6}, 827, 845, 852. 

\noindent Iagolnitzer D (1976a) Physical-Region Properties of Multiparticle 
Collision Amplitudes. In Balian R and
Iagolnitzer D (eds) {\it Structural Analysis of Collision Amplitudes}, pp
161-190. North Holland. 

\noindent Iagolnitzer D (1976b) Analytic Structure of Distributions and 
Essential Support Theory. In Balian R and
Iagolnitzer D (eds) {\it Structural Analysis of Collision Amplitudes}, pp
295-358. North Holland. 

\noindent Iagolnitzer D (1978) {\it The S-Matrix}. North Holland.

\noindent Iagolnitzer D (1981) Analyticity Properties of the S-Matrix: 
Historical Survey and Recent Results in S-Matrix Theory and Axiomatic 
Field Theory. In Mitter H and Pittner L (eds) {\it New Developments in 
Mathematical Physics}, pp 235-328. Springer-Verlag.

\noindent Iagolnitzer D (1993) {\it Scattering in Quantum Field Theories}.
Princeton University. 

\noindent Iagolnitzer D and Stapp H P (1969) {\it Com. Math. Phys.} {\bf 
14}, 15. 

\noindent Landau L D (1955) On the Quantum Theory of
Fields. In Pauli W (ed) {\it Neils Bohr and the Development of Physics},
 pp 52-69 . Pergamon. 

\noindent Landau L D (1959) {\it Nucl. Phys.} {\bf 13}, 181.

\noindent Landau L D (1960) Fundamental Problems. In 
Fierz M and Weisskopf V F (eds) {\it Theoretical Physics in the Twentieth 
Century - A Memorial Volume to Wofgang Pauli}, pp 245-248 . Interscience.

\noindent Landau L D  and Pomeranchuk I Ya (1955) {\it Dokl. Akad. Nauk.
SSSR} {\bf 102}, 489.

\noindent Lehmann H, Symanzik K and Zimmerman W (1957) {\it Nuovo Cim.}
{\bf 6}, 319.

\noindent Low F (1966) Comment. In {\it Proc. XIIIth International 
Conference on High Energy Physics}, p 249. University of California. 

\noindent Low F (1974) Fifty Years of Quantum Field Theory. In Zichichi A 
(ed) {\it Lepton and Hadron Structure}, p 938. Academic Press. 

\noindent Mandelstam S (1958) {\it Phys. Rev.} {\bf 112}, 1344.

\noindent Mandelstam S (1961) Two-Dimensional Representations of Scattering
Amplitudes and their Applications. In Stoops R (ed.) {\it The Quantum Theory 
of Fields, the 12th Solvay Conference}. Interscience. 

\noindent Mandelstam S (1967) Dynamics Based on Indefinitely Rising Regge 
Trajectories. In Hagen C R, Guralnik G and Mathur V A (eds) 
{\it Proc. 1967 International
Conference on Particles and Fields}, pp 604-615. Interscience. 

\noindent Mandelstam S (1976) Dual Resonance Models. In Balian R and
Iagolnitzer D (eds) {\it Structural Analysis of Collision Amplitudes}, pp
593-637. North Holland.

\noindent Migdal A A, Polyakov A M and Ter-Martirosyan K A (1974) 
{\it Zh. Eksp. Teor.  Fiz.} {\bf 67}, 84.

\noindent Moshe M (1978) {\it Phys. Repts.} {\bf 37C}, 255. 

\noindent Mussardo G (1992) {\it Phys. Repts.} {\bf 218}, 215. 

\noindent Olive D (1964) {\it Phys. Rev.} {\bf 135B}, 745. 

\noindent Politzer H D (1974) {\it Phys. Repts.} {\bf 14}, 129.

\noindent Sato M (1975) Recent Developments in Hyperfunction Theory and 
it's Application to Physics. In {\it Lecture Notes in Physics} {\bf 39}.
Springer-Verlag. 

\noindent Stapp H P (1962) {\it Phys. Rev.} {\bf 125}, 2139.

\noindent Stapp H P (1976a) Discontinuity Formulas for Multiparticle 
Amplitudes. In Balian R and
Iagolnitzer D (eds) {\it Structural Analysis of Collision Amplitudes}, pp
191-274. North Holland.

\noindent Stapp H P (1976b) Many-particle Dispersion Relations. 
In Balian R and
Iagolnitzer D (eds) {\it Structural Analysis of Collision Amplitudes}, pp
191-274. North Holland.

\noindent Stapp H P and White A R (1982) {\it Phys. Rev.} {\bf D26}, 2145. 
(1982). 

\noindent 't Hooft G (1980) Confinement and Topology in Non-Abelian Gauge 
Theories. In Urban P (ed) {\it Field Theory and Strong Interactions}, 
pp 531-586, Springer-Verlag.

\noindent 't Hooft G (1999) {\it Nucl. Phys. Proc. Suppl.} {\bf 74}, 413. 

\noindent Weinberg S (1996) {\it The Quantum THeory of Fields, Vol I}.
Cambridge University.

\noindent White A R (1976) The Analytic Foundations of Regge Theory. In 
R.~Balian R and Iagolnitzer D (eds) 
{\it Structural Analysis of Collision Amplitudes}, pp 427-503. North Holland.

\noindent White A R (1991) {\it Int. J. Mod. Phys.} {\bf A11}, 1859. 

\noindent White A R (1993) {\it Int. J. Mod. Phys.} {\bf A8}, 4755.

\noindent White A R (1994) {\it Phys. Lett.} {\bf B334}, 87.

\noindent White A R (1998) {\it Phys. Rev.} {\bf D58}, 074008.

\noindent White A R (1999) The Triangle Anomaly in Triple Regge Limits.
hep-ph/9910458. 

\noindent Zamolodchikov A B and Zamolodchikov Al B (1979) {\it Ann. Phys.}
{\bf 120}, 253. 
}
\end{document}